\documentclass{aa}  

\usepackage{graphicx}
\usepackage{txfonts}
\usepackage{lscape}
\usepackage{hyperref}
\begin{document}

\title{Magnetism, rotation, and nonthermal emission in cool stars\thanks{Table
    B.1 is available in electronic form at the CDS via anonymous ftp to
    cdsarc.u-strasbg.fr (130.79.128.5) or via
    \url{http://cdsweb.u-strasbg.fr/cgi-bin/qcat?J/A+A/}. MCMC posteriors
    together with all spectral line fits and field component distributions are
    available at \url{http://carmenes.cab.inta-csic.es/}.}} \subtitle{Average
  magnetic field measurements in 292 M dwarfs}

\titlerunning{Magnetism, rotation, and non-thermal emission in cool stars}
   
   \author{A. Reiners\inst{1}
          \and
          D. Shulyak\inst{2}
          \and
          P.J. K\"apyl\"a\inst{1}
          \and
          I. Ribas\inst{3, 4}
          \and
          E. Nagel\inst{5}
          \and
          M. Zechmeister\inst{1}
          \and
          J.A. Caballero\inst{6}
          \and
          Y. Shan \inst{1, 7}
          \and \\
          B. Fuhrmeister\inst{5}
          \and
          A. Quirrenbach\inst{8}
          \and
          P.J. Amado\inst{2}
          \and
          D. Montes \inst{9}
          \and
          S.V. Jeffers \inst{10}
          \and
          M. Azzaro \inst{11}
          \and
          V.J.S. B\'ejar \inst{12,13}
          \and \\
          P. Chaturvedi \inst{14}
          \and
          Th. Henning \inst{15}
          \and
          M. K\"urster \inst{15}
          \and
          E. Pall\'e \inst{12,13}
        }
        \institute{
          Institut f\"ur Astrophysik, Georg-August-Universit\"at, D-37077 G\"ottingen, Germany
          \and
          Instituto de Astrof\'isica de Andaluc\'ia (Consejo Superior de Investigaciones Cient\'ificas), E-18008 Granada, Spain
          \and
          Institut de Ci\`encies de l'Espai (Consejo Superior de Investigaciones Cient\'ificas), E-08193 Bellaterra, Barcelona, Spain
          \and
          Institut d'Estudis Espacials de Catalunya, E-08034 Barcelona, Spain
          \and
          Hamburger Sternwarte, Universit\"at Hamburg, D-21029 Hamburg, Germany
          \and
          Centro de Astrobiolog\'ia (Consejo Superior de Investigaciones Cient\'ificas – Instituto
          Nacional de T\'ecnica Aeroespacial), E-28692 Villanueva de la Ca\~nada, Madrid, Spain
          \and
          Centre for Earth Evolution and Dynamics, Department of Geosciences,
          University of Oslo, Sem S{\ae}lands vei 2b 0315 Oslo, Norway
          \and
          Landessternwarte, Zentrum f\"ur Astronomie der Universit\"at Heidelberg,
          D-69117 Heidelberg, Germany
          \and
          Departamento de F\'isica de la Tierra y Astrof\'isica \& Instituto de F\'isica de Part\'iculas y
          del Cosmos, Facultad de Ciencias F\'isicas, Universidad Complutense de Madrid, E-
          28040 Madrid, Spain
          \and
          Max-Planck-Institut für Sonnensystemforschung, D-37077, G\"ottingen, Germany
          \and
          Centro Astron\'omico Hispano-Alem\'an, Observatorio de Calar Alto,
          E-04550 G\'ergal, Almer\'ia, Spain
          \and
          Instituto de Astrof\'isica de Canarias, E-38205 La Laguna, Tenerife, Spain
          \and
          Departamento de Astrof\'isica, Universidad de La Laguna, E-38206 La
          Laguna, Tenerife, Spain
          \and
          Th\"uringer Landessternwarte Tautenburg, D-07778 Tautenburg, Germany
          \and
          Max-Planck-Institut f\"ur Astronomie, D-69117 Heidelberg, Germany
        }
   \date{Received Feb 02, 2022 / Accepted Mar 18, 2022}

   \abstract{Stellar dynamos generate magnetic fields that are of fundamental
     importance to the variability and evolution of Sun-like and low-mass
     stars, and for the development of their planetary systems. As a key to
     understanding stellar dynamos, empirical relations between stellar
     parameters and magnetic fields are required for comparison to ab initio
     predictions from dynamo models. We report measurements of surface-average
     magnetic fields in 292 M dwarfs from a comparison with radiative transfer
     calculations; for 260 of them, this is the first measurement of this
     kind. Our data were obtained from more than 15,000 high-resolution
     spectra taken during the CARMENES project. They reveal a relation between
     average field strength, $\langle B \rangle$, and Rossby number, $Ro$,
     resembling the well-studied rotation-activity relation. Among the slowly
     rotating stars, we find that magnetic flux, $\Phi_\textrm{B}$, is
     proportional to rotation period, $P$, and among the rapidly rotating
     stars that average surface fields do not grow significantly beyond the
     level set by the available kinetic energy. Furthermore, we find close
     relations between nonthermal coronal X-ray emission, chromospheric
     H$\alpha$ and Ca~H\&K emission, and magnetic flux. Taken together, these
     relations demonstrate empirically that the rotation-activity relation can
     be traced back to a dependence of the magnetic dynamo on rotation.  We
     advocate the picture that the magnetic dynamo generates magnetic flux on
     the stellar surface proportional to rotation rate with a saturation limit
     set by the available kinetic energy, and we provide relations for average
     field strengths and nonthermal emission that are independent of the
     choice of the convective turnover time. We also find that Ca\,H\&K
     emission saturates at average field strengths of
     $\langle B \rangle \approx 800$\,G while H$\alpha$ and X-ray emission
     grow further with stronger fields in the more rapidly rotating
     stars. This is in conflict with the coronal stripping scenario
     predicting that in the most rapidly rotating stars coronal plasma would
     be cooled to chromospheric temperatures. }
% 5 {} token are mandatory

%   \abstract{}
   

   \maketitle
%
%-------------------------------------------------------------------

\section{Introduction}

Sun-like and low-mass stars generate magnetic fields through a hydromagnetic
dynamo operating in their interiors \citep{1955ApJ...122..293P,
  2013SAAS...39.....C}. The stellar dynamo is believed to transform the
kinetic energy, $E_\textrm{kin}$, of the star's turbulent convective motion
into magnetic energy, $E_\textrm{B}$. In analogy to the Sun, starspots and
stellar activity are believed to be a consequence of the emergence of magnetic
flux at the stellar surface \citep[e.g.,][]{2015Sci...347.1333C,
  2017LRSP...14....4B}. Heating in the stellar corona and chromosphere is
closely related to magnetic flux \citep{2003ApJ...598.1387P,
  2004A&ARv..12...71G}, and magnetically active regions cause most of the
stellar energy output variation on timescales ranging from minutes to centuries
\citep{2004A&ARv..12..273F}.

Magnetic fields determine the variability and rotational evolution of
main-sequence stars as well as the evolution of planetary systems. The
relation between angular momentum loss and age yields a predictable evolution
of stellar rotation and can provide estimates for stellar age from rotation
measurements \citep{1972ApJ...171..565S}, a method called gyrochronology
\citep{2003ApJ...586..464B}. This spin-down mechanism acts most effectively in
young stars and, over time, transforms them into old field stars with terminal
rotation rates that are mass-dependent \citep{2014ApJS..211...24M}.  Magnetic
activity influences planetary evolution through the amount of high-energy
radiation emitted during the phase when the planetary atmosphere is young
because intense high-energy radiation and winds can evaporate the atmosphere
of a planet in a close orbit \citep{2011A&A...532A...6S,
  2015ApJ...815L..12J}. Magnetic fields and their variability are among the
main obstacles for the radial velocity detection of low-mass planets
\citep{2013A&A...552A.103R, 2020MNRAS.497.4009L, 2020arXiv200513386H,
  2021arXiv210714291C}.

It is often suggested that the well-established stellar activity-rotation
relation \citep[e.g.,][]{1972ApJ...171..565S, 1984ApJ...279..763N,
  2003A&A...397..147P, 2011ApJ...743...48W, 2018MNRAS.479.2351W, 2014ApJ...794..144R} is a causal
consequence of a hydromagnetic dynamo generating stronger magnetic fields in
more rapidly rotating stars. The influence of rotation on convective motion is
often expressed in terms of the Rossby number, $Ro = P/\tau$, the ratio
between rotation period, $P$, and convective turnover time, $\tau$;
chromospheric and coronal emission are observed to saturate in stars rotating
more rapidly than $Ro \approx 0.1$. Slowly rotating stars ($Ro \ga 0.1$) emit
chromospheric and coronal emission in approximate proportion to the inverse of
the rotational period squared, $P^{-2}$, which leads to a weakening of stellar
activity with age because of angular momentum loss. A mild dependence between
X-ray emission and rotation is also observed in the saturated regime
\citep{2003A&A...397..147P, 2014ApJ...794..144R, 2020A&A...638A..20M}.

There is some direct evidence of the magnetic dynamo showing a similar
dependence on rotation; the magnetic energy generated in rapidly rotating
planets and stars is approximately limited by kinetic energy flux from
convection \citep{2009Natur.457..167C}, and evidence for saturation in
magnetic field strength exists from semi-empirical methods
\citep{2009ApJ...692..538R} and from detailed modeling in a limited sample
\citep{2017NatAs...1E.184S}. Furthermore, observations in very active M dwarfs
reveal a relation between the field strength and rotation rate
\citep{2017NatAs...1E.184S, 2019A&A...626A..86S, 2021A&ARv..29....1K}, and
observations of large-scale surface magnetic fields from Zeeman Doppler
imaging (ZDI) show a similar relation \citep{2014MNRAS.441.2361V}.

A major challenge in predicting stellar magnetic field strengths is the broad
range of timescales and length-scales involved, rendering detailed simulations of
stellar convection impractical \citep{2017LRCA....3....1K}. Instead,
simplified models approximate the effects of turbulent convection, rotation,
and Lorentz force feedback \citep{2017LRSP...14....4B}. For example, a balance
between Coriolis, buoyancy, and Lorentz forces (MAC balance) leads to
$E_\textrm{B} \sim E_\textrm{kin}/Ro$. On the other hand, a balance between
advection and Lorentz forces results in magnetic fields in equipartition with
convective kinetic energy. Scaling relations between magnetic dynamo
efficiency and luminosity or rotation rate can in principle be estimated from
models \citep[e.g.,][]{2019ApJ...876...83A}, but the extent to which such
simplified models can be compared to physical objects remains
unclear. Furthermore, direct magnetic field measurements and the range of
stellar parameters covered by observations have so far not provided enough
empirical evidence to guide dynamo models, especially in slowly rotating
stars.

The signatures of magnetism on surfaces of Sun-like and low-mass stars are
very subtle. The most direct diagnostic of stellar magnetic fields is the
Zeeman effect \citep[e.g.,][]{2004ASSL..307.....L, 2009ARA&A..47..333D,
  2012LRSP....9....1R, 2021A&ARv..29....1K}. In general, it can be observed in
polarized or in unpolarized light. Measurements in polarized light can detect
the distribution of very weak fields on the order of a few Gauss. The analysis
of circularly polarized light alone is prone to cancellation effects, but the
measurement of linearly polarized light is very demanding. Often,
deconvolution techniques are employed to construct average line profiles with
very high signal-to-noise ratios (S/N) \citep[e.g.,][]{1989A&A...225..456S,
  1993A&A...278..231S, 2015ApJ...805..169R}.  Unpolarized light can
potentially reveal the full and unbiased magnetic field including very
small-scale components, but here averaging spectral lines to boost the signal
is not possible, and line formation must be modeled in great detail
\citep[see][]{2021A&ARv..29....1K}. Direct measurement of the surface-average
magnetic field, $\langle B \rangle$, therefore requires exceptionally high S/N
(at high spectral resolution) plus sophisticated radiative transfer
calculations that can model polarization. In consequence, only a limited
number of average field measurements based on detailed profile modeling exist,
and no empirical relation is known between magnetic fields, fundamental
stellar parameters, and rotation. On the theory side, ab initio predictions
about stellar magnetic fields are very challenging
\citep{2017LRSP...14....4B}, and empirical information about magnetic fields
could help identify the parameter space in which stellar dynamos can operate.

Collecting data that meet the high requirements for average magnetic field
measurements in a sizeable sample of low-mass stars is a challenge that can
hardly be met by programs investigating the stars or stellar activity
alone. On the other hand, radial velocity surveys searching for planetary
companions around low-mass stars acquire extensive data sets that can also be
used for the study of the host stars. The data we present in this paper were
collected during the course of the CARMENES survey for exoplanets around M
dwarfs \citep{2018A&A...612A..49R, 2020SPIE11447E..3CQ}. Average magnetic
field measurements in a subsample of very active stars were already presented
in \citet{2019A&A...626A..86S}. Here, we investigate data from the full
CARMENES sample and present average magnetic fields for active and inactive
stars.

\section{Data}

For the CARMENES survey, we observed more than 300 M dwarfs since early 2016
with the goal to monitor radial velocities, leading to a number of exoplanet
discoveries \citep[e.g.,][]{2018Natur.563..365R, 2019Sci...365.1441M,
  2019A&A...627A..49Z}. The sample used here is based on the one from
\citet{2018A&A...612A..49R} and includes a number of stars that were included
later. We excluded the multiple systems reported in
\citet{2021A&A...653A..49B} as well as visual binaries. CARMENES is operating
at the 3.5m telescope at Calar Alto observatory, Spain, and consists of the
two channels VIS and NIR that cover wavelength ranges 5200--9600\,\AA\ (VIS)
and 9600--17,100\,\AA\ (NIR) at spectral resolution
$R = \lambda / \Delta \lambda$ of 94,600 and 80,400, respectively
\citep{2016SPIE.9908E..12Q}. Data from both channels were used for this work.

Data were reduced with the \texttt{caracal} pipeline
\citep{2016SPIE.9910E..0EC} using optimal extraction
\citep{2014A&A...561A..59Z}. We computed radial velocities for each spectrum
and co-added individual spectra to obtain a master spectrum for each star
using the \texttt{serval} package \citep{2018A&A...609A..12Z}. Our master
spectrum is therefore a time-average of many spectra obtained during the
CARMENES survey. Before co-addition, we modeled and removed atmospheric
absorption lines from the Earth's atmosphere with the package
\texttt{molecfit} \citep{Smette15} as described in \citet{NagelPhD}. The S/N
of the final spectra depends on the number of observations per star and the
amount of telluric contamination. Typical numbers of individual exposures per
star are between 10 and 100, and typical values of S/N are in the 100--1000
range. The sample of stars used for our analysis, the number of spectra
co-added for each star, and the approximate S/N around $\lambda = 8700$\,\AA\
are provided in Table\,B.1.

\section{Analysis}
\label{Sect:Analysis}

\begin{table}
  \centering
  \caption{\label{tab:lines}Absorption lines used for our analysis. The
    tabulated wavelengths are valid for vacuum.}
    \begin{tabular}{crc}
      \hline
      \hline
      \noalign{\smallskip}
      Species & $\lambda$ (\AA) & Land\'e $g$ \\
      \noalign{\smallskip}
      \hline
      \noalign{\smallskip}
      \ion{Ti}{I} & 8355.46  & 2.25 \\
      \ion{Ti}{I} & 8399.21  & 0.00 \\ 
      \ion{Ti}{I} & 8414.67  & 0.66 \\
      \ion{Fe}{I} & 8470.73  & 2.49 \\
      \ion{Fe}{I} & 8516.41  & 1.83 \\
      \ion{Fe}{I} & 8691.01  & 1.66 \\
      \ion{Ti}{I} & 9678.20  & 1.35 \\
      \ion{Ti}{I} & 9691.53  & 1.50 \\
      \ion{Ti}{I} & 9731.07  & 1.00 \\
      \ion{Ti}{I} & 9746.28  & 0.00 \\
      \ion{Ti}{I} & 9786.13  & 1.48 \\
      \ion{Ti}{I} & 9790.37  & 1.50 \\
      \ion{FeH}{} & 9957.02  &      \\
      \ion{K}{I}  & 12435.68 & 1.33 \\
      \ion{K}{I}  & 12525.56 & 1.17 \\
      \noalign{\smallskip}
      \hline
      \noalign{\smallskip}
    \end{tabular}
\end{table}

We measured average magnetic field strength $\langle B \rangle$ from
comparison of spectral absorption lines to radiative transfer calculations as
explained in \citet{2017NatAs...1E.184S, 2019A&A...626A..86S}. As demonstrated
there, polarized radiative transfer calculations for atmospheres of M-type
stars can reproduce observed line profiles with relatively high quality. The
degeneracy between line broadening mechanisms can be overcome using lines with
different Zeeman sensitivities, that is, including lines with low and high
Land\'e-g factors simultaneously. Furthermore, several lines in the
near-infrared part of the spectrum (8000--10,000\,\AA) are relatively strong
and show potentially observable Zeeman enhancement caused by ``desaturation''
through the wavelength shifts of the individual absorption components in the
presence of a magnetic field \citep{1992ApJ...390..622B}.

\subsection{Line selection}

The selection of spectral lines is crucial for Zeeman broadening analysis, we
refer to \citet{2021A&ARv..29....1K} for a detailed review. The set of lines
should cover a range of Land\'e-g values to break degeneracies between
different broadening mechanisms. Lines at longer wavelengths are more Zeeman
sensitive and therefore preferable \citep{2009ARA&A..47..333D,
  2012LRSP....9....1R}. Unfortunately, the density of absorption lines in M
dwarf spectra is relatively low beyond 10,000\,\AA, and lines are often
severely contaminated by telluric absorption. \citet{2017ApJ...835L...4K} and
\citet{2017NatAs...1E.184S} showed that a group of \ion{Ti}{i} lines around
9700\,\AA\ is very useful for Zeeman analysis, and some lines of molecular FeH
that are relatively free of telluric contamination are available around
10,000\,\AA. We identified a set of suitable spectral lines with a range of
Land\'e-g values to separate magnetic Zeeman broadening from other broadening
effects. Spectral lines available for comparison are summarized in
Table\,\ref{tab:lines}. The list includes atomic lines from \ion{Ti}{i},
\ion{Fe}{i}, \ion{K}{i}, and a particularly useful pair of lines from
molecular FeH. Except for the two \ion{K}{i} lines, we could not use any other
line beyond $\lambda = 10,000$\,\AA.

For radiative transfer calculations, we used stellar atmosphere models from
the MARCS library \citep{2008A&A...486..951G}. Based on these models, we
computed synthetic spectra on a grid of effective temperatures and surface
gravity with step sizes of$\Delta T_{\rm eff} = 100$\,K in the
$T_{\rm eff}$ = 2500--4000\,K range, and $\log{g}$ = 4.5, 5.0, and 5.5 dex. For each
star, we linearly interpolated the specific synthetic spectra according to
their value of $T_{\rm eff}$ and $\log{g}$ from the grid (see
Sect.\,\ref{sect:sample} and Table\,B.1; we computed $\log{g}$
from mass and radius). The required data about atomic absorption lines,
including Land\'e-g factors, were taken from VALD line lists
\citep{1995A&AS..112..525P, 1999A&AS..138..119K}. For lines from the molecule
FeH, we followed a semi-empirical approach to compute the wavelength-shifting
of individual Zeeman components \citep{2008A&A...482..387A,
  2010A&A...523A..37S}.

For each star, we selected a subset of lines from the available line list
based on the temperature of the star and the quality of the observed data in
the spectral range of each line. A line was selected only if the average S/N
in its wavelength range exceeded a value of 50. Some lines are located in
close vicinity of telluric absorption lines. Depending on the radial velocity
of the star and the time of individual observations, these lines could be more
or less affected by telluric contamination and could be selected in some stars
but not in all. The intensity of the lines and also of line blends are strong
functions of stellar temperature. We investigated the usefulness of the lines
from our list as a function of stellar temperature and employed the following
scheme. For stars with $T_{\rm eff} > 3750$\,K, we used all the available
lines from Table\,\ref{tab:lines}. In stars cooler than
$T_{\rm eff} = 3750$\,K, the lines at $\lambda$ = 8355.46 and 8470.73\,\AA\
were not included in the fit, and in stars cooler than
$T_{\rm eff} = 3600$\,K, the line $\lambda = 8516.41$\,\AA\ was also excluded. In addition to atomic \ion{Ti}{I} and \ion{Fe}{I} lines, we included the
\ion{K}{I} lines at $\lambda = 12435.68$ and 12525.56\,\AA\ in our list. These
lines carry relevant information in the presence of strong magnetic fields
\citep[$B \ga 1$\,kG;][]{2022A&A...657A.125F} but are less useful in weakly
magnetic stars because of the strong intrinsic line broadening. We therefore
included the \ion{K}{I} lines only in very active stars with
$\log{L_{\textrm{H}\alpha}/L_{\rm bol}} > -4.6$. All fits were visually
inspected, and we removed individual lines in cases where the spectra were
obviously affected by systematics. This was necessary in cases where
contamination by telluric lines removed a substantial part of the line profile
while our threshold average S/N of 50 was still met. In a few stars,
individual orders of our spectra were affected by systematics from template
construction and/or telluric correction. The multiline approach can only reliably
disentangle Zeeman broadening from rotational broadening if the lines
cover a range of different Land\'e-g values. It is particularly important to
include at least one line that shows little or no sensitivity to Zeeman
broadening in order to determine $\varv\sin{i}$ and disentangle rotational
from Zeeman broadening. For all magnetic field strength measurements included
in Table\,B.1, four or more spectral lines were used, of which at
least one line has Land\'e-$g = 0.0$.

\subsection{Fitting strategy}

\begin{figure*}
  \centering
  \mbox{
    \parbox{.95\textwidth}{
      \resizebox{.95\textwidth}{!}{\includegraphics[viewport=20 102 720 289, clip=]{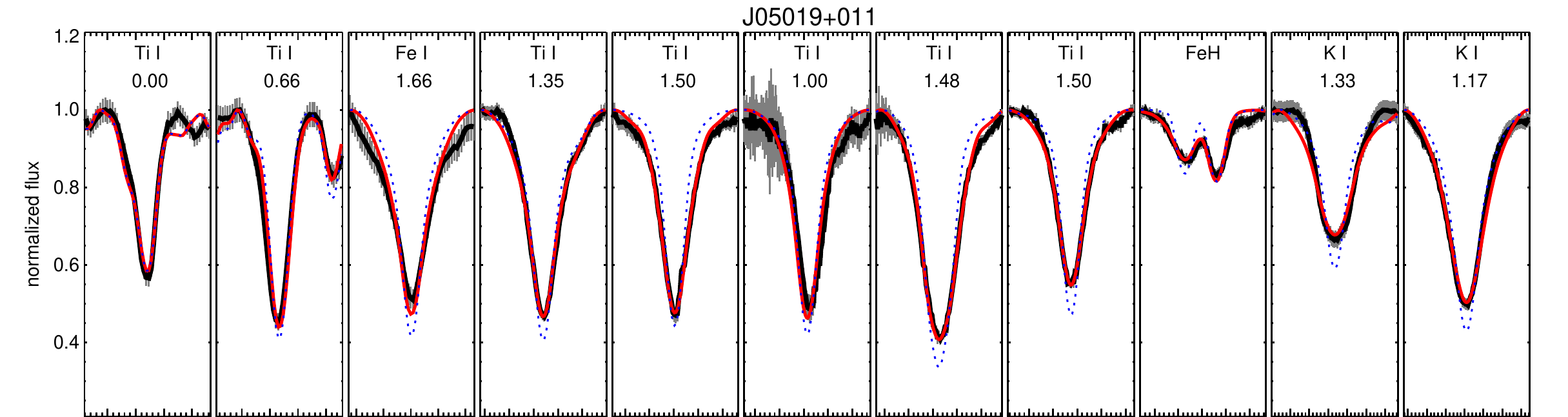}}\\
      \resizebox{.95\textwidth}{!}{\includegraphics[viewport= 20 5 720  27, clip=]{linefit_J05019+011.eps}}
    }}
  \resizebox{\textwidth}{!}{\includegraphics{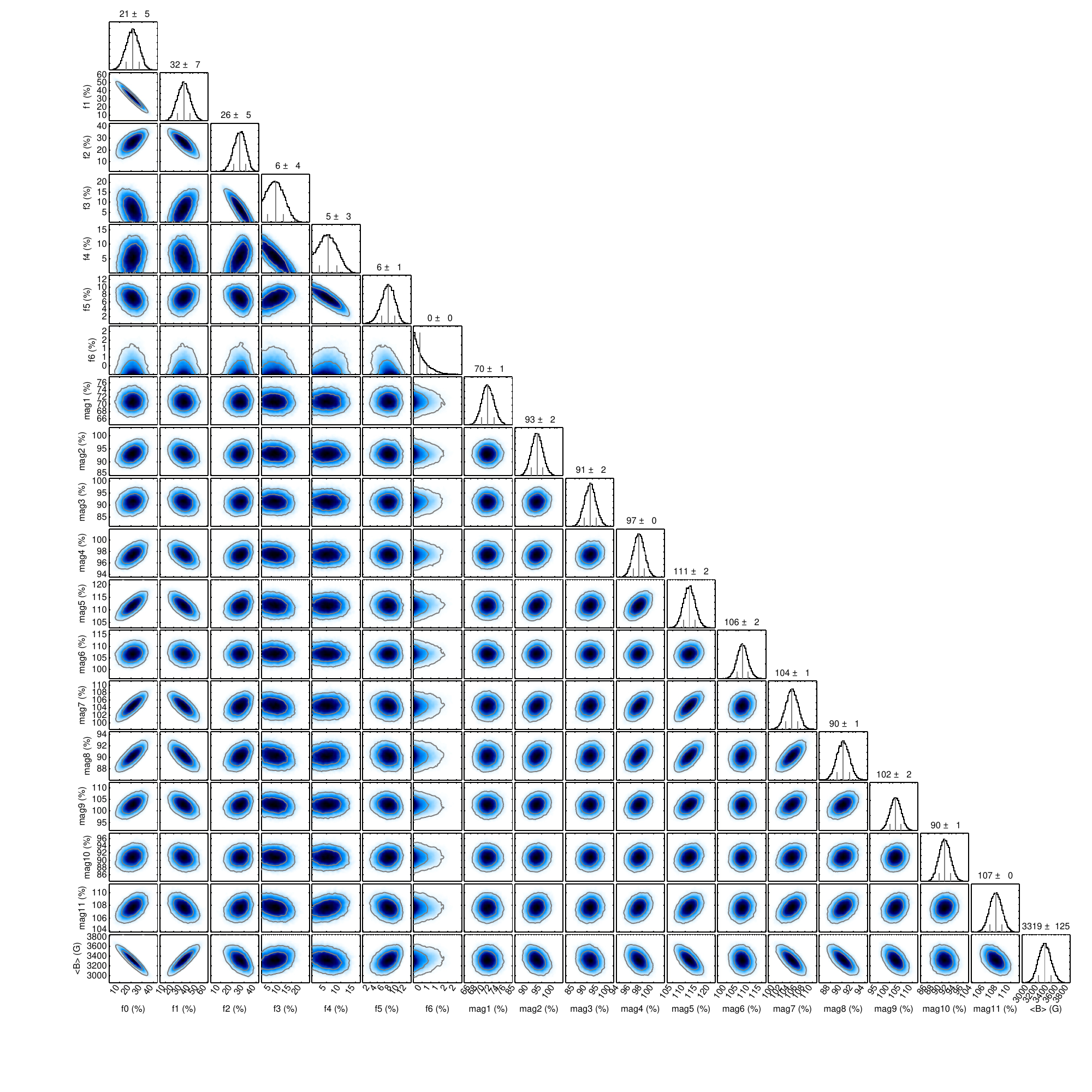}}\\[-18cm]
  \hspace{.55\textwidth} \resizebox{0.4\textwidth}{!}{\includegraphics[viewport= 30 5 640 469, clip=]{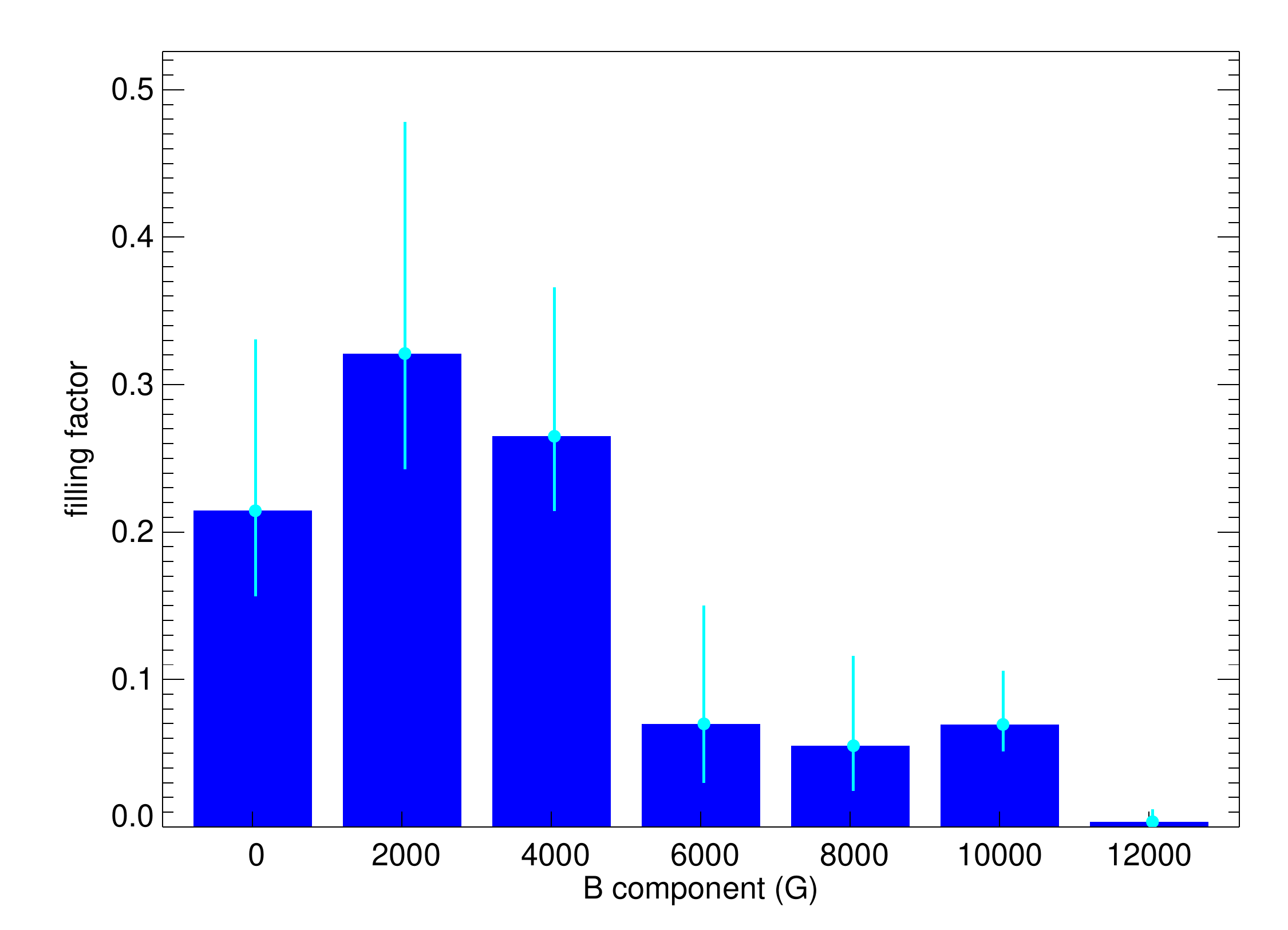}}\\[11.5cm]
  \caption{\label{fig:ExampleFit}\emph{Top panel:}Example fit for one of our
    stars; J05019+011 (M4.0\,V). Data are shown as black line with
    uncertainties. The best fit solution is shown as red line, the best
    solution with $\langle B
    \rangle$$=0$\,G as blue dashed line. The absorbing species are indicated
    together with the effective Land\'e-g factor for atomic
    lines. \emph{Bottom panel:} Cornerplot for posterior MCMC parameter
    distribution. In this example, six field components
    $f_1$--$f_6$ were used, and twelve lines were fit for which one scaling
    parameter
    (mag$_1$--mag$_{11}$) per line is shown. Corresponding lines are shown in
    the upper panel in the same order. The posterior distribution for $\langle
    B \rangle=\Sigma B_i
    f_i$ is included in the plot at the bottom row. \emph{Inset panel:}
    Relative distribution of filling factors $f_i$ for field components
    $B_i$. Uncertainties of individual field components are indicated as cyan
    lines.}
\end{figure*}

We observed that simultaneously fitting
$\varv\,\sin{i}$ and Zeeman broadening often leads to overestimated rotational
broadening, which is visible when the model predicts profiles that are too
broad in the Land\'e-$g =
0.0$ lines. Therefore, we first determined
$\varv\,\sin{i}$ using only the Land\'e-$g =
0.0$ lines. We emphasize that rotation significantly smaller than instrumental
broadening ($\varv\,\sin{i} \la
$1--2\,km\,s$^{-1}$) has little effect on the line profile. In a second step,
we fixed
$\varv\,\sin{i}$ and determined the magnetic field distribution as described
below. A comparison between observations and model spectra was carried out
through computation of
$\chi^2$, that is, the quadratic sum over their residuals. The fitting process
was performed with the MCMC-toolkit SoBAt \citep{2021ApJS..252...11A}. For
each star, we constructed a grid of synthetic line profiles for a predefined
range of magnetic field strengths. The line profiles were modeled from this
grid as the weighted sum of
$B$-components with different field strengths; we divided the surface of the
star into spatial components that each contribute to the total spectrum with a
local spectrum according to the field strength chosen for that
$B$ component. The weight of each
$B$ component is defined by its filling factor. The choice of field components
available to the fit procedure was made according to the predetermined
$\varv\,\sin{i}$ and the
H$\alpha$ emission of the star. Because the Zeeman shift of our most
magnetically sensitive lines is approximately
2.5\,km\,s$^{-1}$\,kG$^{-1}$, we sampled the field distribution in steps of
$\Delta B =
1$\,kG for stars with a projected rotational velocity $\varv\sin{i} < 5
$km\,s$^{-1}$ and $\Delta B =
2$\,kG for stars with higher rotational broadening. This strategy should
minimize degeneracies between the individual field components but capture all
information from the line profiles. Our models included field components in
the range 0--4\,kG in inactive or moderately active stars with
$\log{L_{\textrm{H}\alpha}/L_{\rm bol}} <
-4.5$. For more active stars, components up to 12\,kG were used. We confirmed
that the choice of field components in number and range did not significantly
influence the results. Our fitting strategy optimized the fit for $N_{\rm c} +
N_{\rm l} - 1$ free parameters, with $N_{\rm
  c}$ the number of magnetic field components and $N_{\rm
  l}$ the number of lines. The free parameters were the $N_{\rm c} -
1$ weights ($f_{i}$) for the individual field components $i$, with $\Sigma_{i}
f_{i} = 1$, and $N_{\rm
  l}$ line scaling parameters, one for each line. With the scaling parameters,
we made it possible to adjust each line strength according to an optical depth
scaling law. The scaling effectively compensates for uncertainties in
oscillator strengths and element abundance. We used uninformative priors for
all free parameters in the [0, 1] range for the field components' weights and
in the [0.3, 1.3] range for the line strength scaling.

\subsection{Parameter uncertainties}

Estimating uncertainties from high-S/N line fits is notoriously difficult,
although an MCMC method provides a convenient way to estimate and visualize
uncertainties and also degeneracies between free parameters. As uncertainties
of our magnetic field measurements, we report 2$\sigma$ uncertainties from the
MCMC distributions. The underlying assumption is that residuals between a
model and observations ($\chi^2$) are caused by statistical processes but not
by a systematic model mismatch. In our spectra, however, the photon noise is
often far smaller than systematic uncertainties expected in our data that are
caused, for example, by limited precision in normalization and co-adding and
by systematic imperfections of the model. Therefore, the photon noise is often
not a good estimator for the likelihood of a fit. Methods that compare
likelihoods can partially overcome and reliably identify the most likely
solution, but parameter uncertainties remain affected by systematic components
in the residuals \citep[see, e.g.,][]{Bonamente17}. We therefore implemented a
two-step procedure to estimate uncertainties. First, we carried out the
fitting procedure with formal photon uncertainties. Then, we multiplied the
photon noise of our data by the square root of $\chi^2_{\nu}$, the reduced
chisquare, calculated for each spectral line, and we carried out the fit
procedure again with the modified uncertainties \citep[in other words, we
assume that our best fit is also a good fit; see][]{1986nras.book.....P}. We
confirmed that the second step did not significantly alter the result but
produces more realistic estimates of our measurement uncertainties. We treated
field measurements as upper limits instead of detections if the result is
consistent within an average field of $\langle B \rangle$ = 100\,G within
2$\sigma$ uncertainties. In these cases, we report the 2$\sigma$ upper limit
in our plots and Table\,B.1. Stars for which only upper limits could be
determined were not considered in our regression curve calculations in the
following.

\subsection{Example and literature comparison}

\begin{figure}
  \centering
  \resizebox{\hsize}{!}{\includegraphics[viewport=0 0 500 455, clip=]{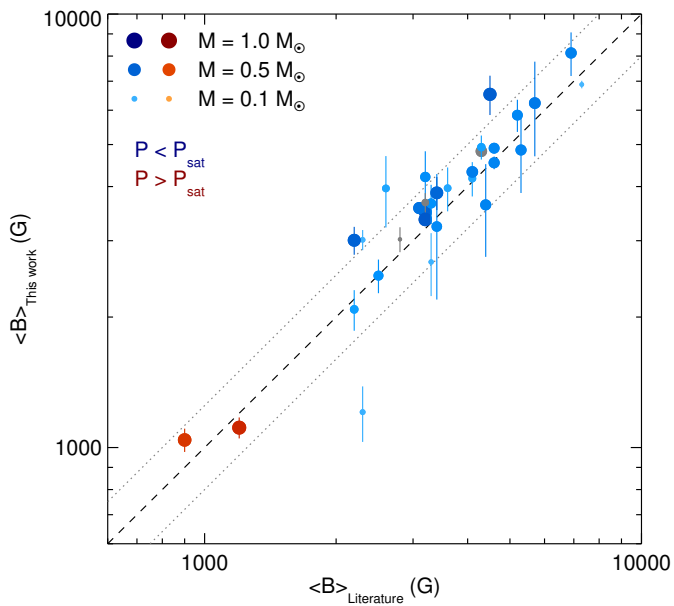}}
  \caption{\label{fig:Literature}Comparison of our average field measurements
    to literature values. Red and blue symbols indicate slow and fast
    rotators, respectively (see Sect.\,\ref{sect:rotact}), and gray symbols show
    stars with no information on rotation period. Symbol size and brightness
    indicate stellar mass. The dashed black line shows identity between our
    measurements and literature values, and the gray dotted lines mark the
    region where the discrepancy is within $\pm\,25$\,\%. }
\end{figure}

We show an example for our line fits in Fig.\,\ref{fig:ExampleFit}. Plots of
all fits are available in electronic format at
\url{http://carmenes.cab.inta-csic.es/}. The distribution of the field
component posteriors shows that there is a degeneracy between components from
adjacent bins, $f_i$ and $f_{i\pm1}$, but there is little crosstalk between
components with very different field strengths. The figures show the observed
data together with the models for all lines and also the distribution of field
components and their uncertainties. We measured the average magnetic field
strength in 292 stars. For 36 of these, the field strength was measured before
by \citet{2017NatAs...1E.184S, 2019A&A...626A..86S} and
\citet{2020ApJ...902...43K}. \citet{2019A&A...626A..86S} also used CARMENES
data focusing on very active stars. The other works are based on data from
other sources. We compare the results of our analysis to those previously
reported in Fig.\,\ref{fig:Literature}. The comparison shows that, except for
a handful of stars in which our new results are up to 50\,\% smaller or larger
than earlier field estimates, the results are typically consistent within
25\,\%. Discrepancies between our and earlier measurements are smaller than
$2\sigma$ in the majority of stars. The main differences between our and
earlier measurements are the selection of lines and the setup of the
model. This demonstrates that formal (statistical) fit uncertainties are often
smaller than systematic uncertainties. We estimate that the accuracy of our
average magnetic field measurements is better than 25\%.

For many of our target stars, \citet{2017MNRAS.472.4563M} reported average
magnetic field measurements from an ``indirect'' method
\citep{2021A&ARv..29....1K}. We provide a comparison between their results and
our values in Appendix\,\ref{sect:Moutou}.

\section{Results}

In this section, we investigate relations between average magnetic fields,
rotation, and nonthermal emission. We augment our sample with stars for which
average magnetic field measurements have been reported in the literature.

\subsection{Sample}
\label{sect:sample}

\begin{figure*}
  \centering
  \mbox{
    \resizebox{.32\hsize}{!}{\includegraphics[viewport=30 5 630 455, clip=]{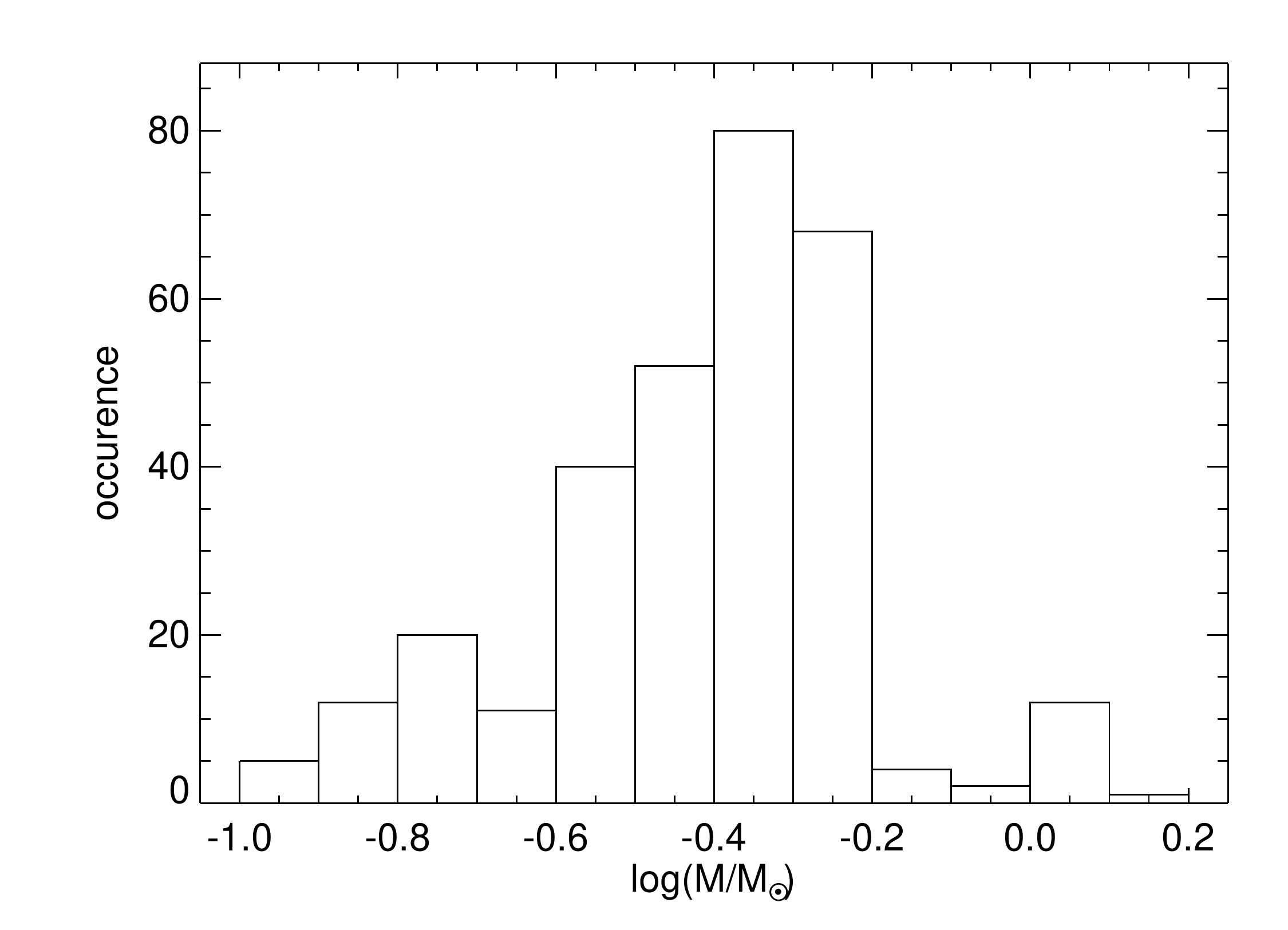}}
    \resizebox{.32\hsize}{!}{\includegraphics[viewport=30 5 630 455, clip=]{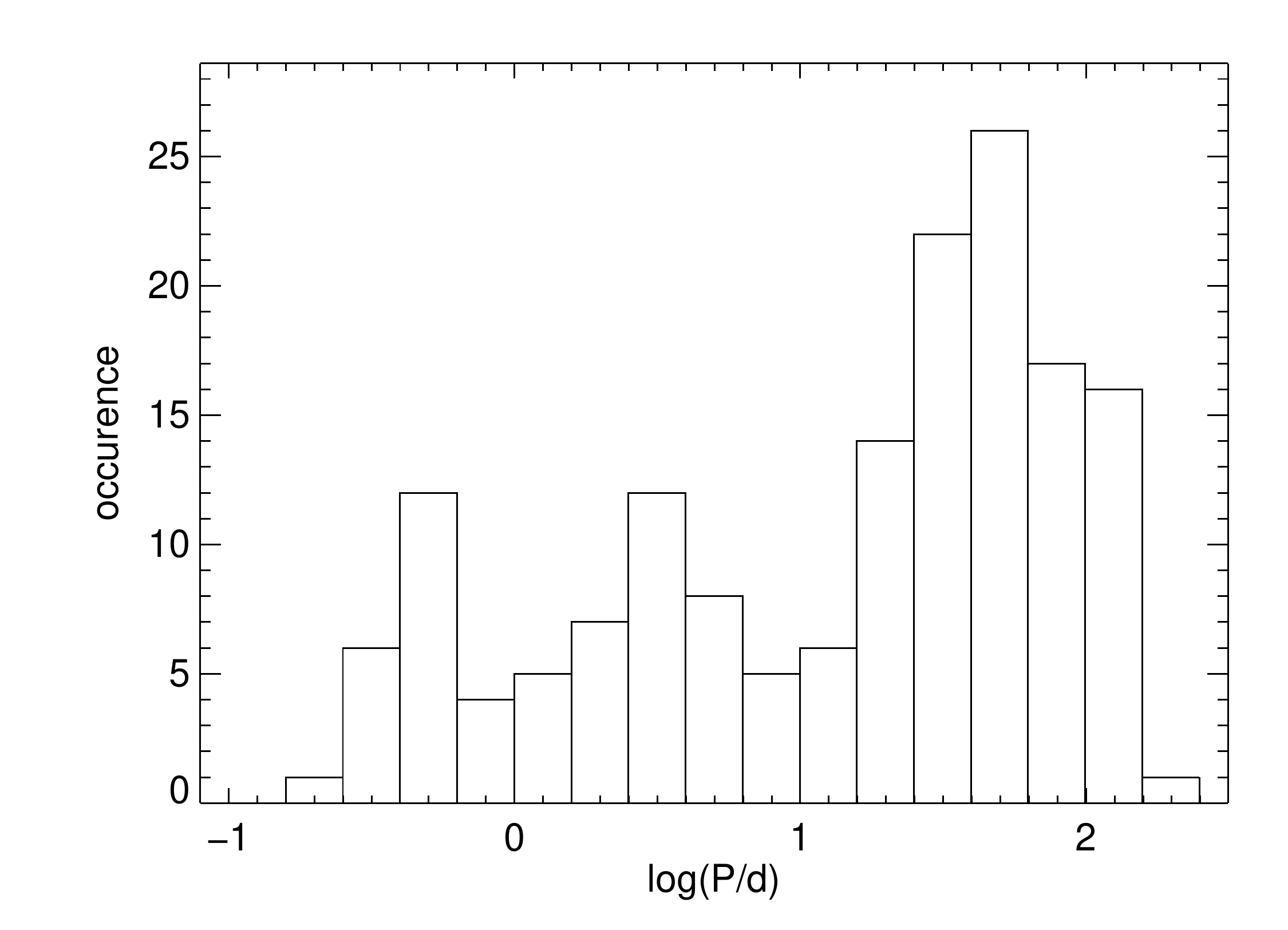}}
    \resizebox{.32\hsize}{!}{\includegraphics[viewport=30 5 630 455, clip=]{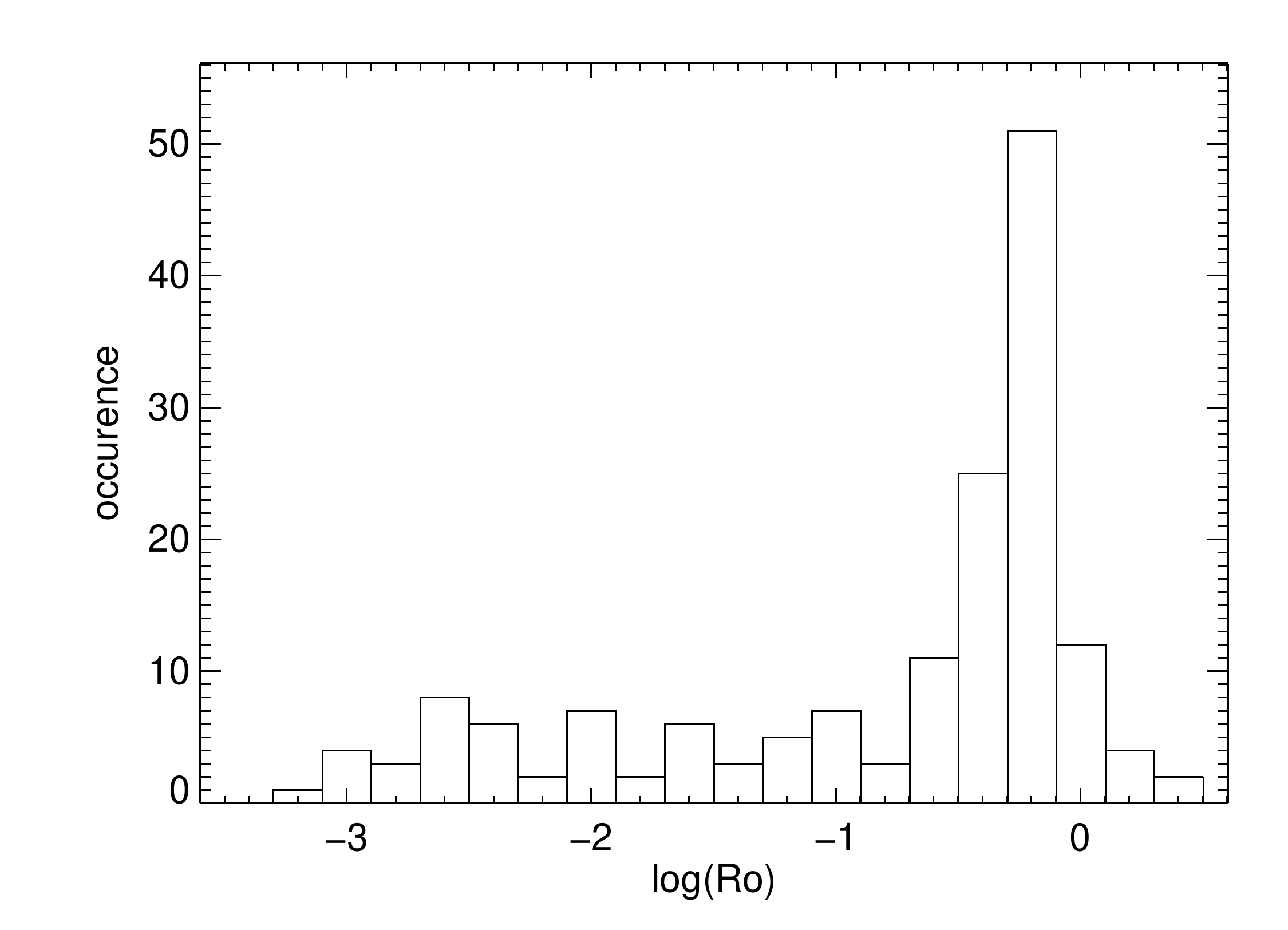}}}
  \caption{\label{fig:Histogram}Distributions of our sample of stars with
    average magnetic field measurements. \emph{Left:} Stellar mass.    \emph{Center:} Rotational period. \emph{Right:} Rossby number. Rotational
    periods are known for 162 of the 314 sample stars, and the remaining 152 stars
    do not appear in the distributions of $P$ and $Ro$.}
\end{figure*}

Our sample on Sun-like and low-mass stars consists of our measurements of 292
M dwarfs observed with CARMENES and 22 additional stars taken from the
literature. From our original sample, we excluded those that did not meet the
quality criteria for the selection of lines as described in
Section\,\ref{Sect:Analysis}. From the literature, we added only magnetic
field measurements from radiative transfer calculations in multiple lines and
considering multiple magnetic field components; following
\citet{2021A&ARv..29....1K}, we added six M dwarfs from
\citet{2017NatAs...1E.184S}. We also included the results on 15 young Sun-like
stars from \citet{2020A&A...635A.142K} and the value for the Sun from
\citet{2004Natur.430..326T}.

The parameters for all 314 stars of our sample are listed in
Table\,B.1. For the CARMENES stars, mass, radius, and luminosity
are taken from \citet{2019A&A...625A..68S}. These values are derived from
PARSEC isochrones and PHOENIX-ACES model fits to the high-resolution spectra
\citep[see][]{2018A&A...615A...6P}. X-ray luminosities are computed from X-ray
fluxes from \citet{1999A&A...349..389V} using distances from
\citet{2021A&A...649A...1G}.  Normalized H$\alpha$ luminosities are from
CARMENES data (see \citet{2019A&A...623A..44S}). They are only reported for the
stars with H$\alpha$ in emission. Ca~H\&K luminosities are estimated from
normalized Ca emission $R'_\textrm{HK}$ in \citet{2021A&A...652A.116P}, we use
$L_\textrm{Ca} = R'_\textrm{HK} L_\textrm{bol}$. The main sources for
uncertainties in X-ray, H$\alpha$, and Ca~H\&K luminosities are variability of
the emission and uncertainties in the stellar parameters used to compute the
luminosities from line equivalent widths. Variability and uncertainties are
typically on the order of a few tenths of a dex \citep[see,
e.g.,][]{2019A&A...623A..44S, 2021A&A...652A.116P}. References for rotation
periods are included in Table\,B.1.  We discarded several
periods measured by \citet{2019A&A...621A.126D} following a reanalysis of the
same photometric data as used in that work but using more conservative
thresholds for significance. As discussed in the references, it cannot be
excluded that a few of the periods are false positives or harmonics of the
real rotation periods. Obvious suspects could potentially be identified in
cases where reported periods are in strong disagreement with the average
relations between rotation and other information. For the young Suns from
\citet{2020A&A...635A.142K}, we collected literature values for mass and
radius from \citet{2007ApJS..168..297T} and \citet{2016AJ....151...59C}.  We
estimate the Rossby number, $Ro = P/\tau$, with $\tau =
12.3$\,d\,$ \times \left(L_\textrm{bol}/L_{\odot}\right)^{-1/2}$, with
rotation period, $P$, in days and bolometric luminosity, $L_\textrm{bol}$. The
expression for $\tau$ was taken from Eq.\,(10) in \citet{2014ApJ...794..144R},
$P_\textrm{sat} = 1.6\,\textrm{d} \times
\left(L_\textrm{bol}/L_{\odot}\right)^{-1/2}$, and
$\tau = P_\textrm{sat}/Ro_\textrm{sat}$. We use $Ro_\textrm{sat} = 0.13$ for
the transition between saturated and non-saturated activity
\citep[see][]{2011ApJ...743...48W, 2018MNRAS.479.2351W, 2014ApJ...794..144R}.
We note that \citet{2018MNRAS.479.2351W} derived a relationship $\tau(M)$ from
X-ray observations in partially and fully convective stars that results in a
similar scaling as the one we use here, but that yields smaller values of
$\tau$ particularly for very low stellar mass. We confirmed that the main
results of our analysis remain valid for $\tau(M)$ as expressed in
\citet{2018MNRAS.479.2351W}, and we emphasize that the relations we report in
the following are independent of the choice of $\tau$.

The distribution of mass, $M$, rotation period, $P$, and Rossby number are
shown in Fig.\,\ref{fig:Histogram}. The sample covers one order of magnitude
in stellar mass with a concentration in the mass range
$M$\,=\,0.4--0.5\,M$_{\odot}$. The original motivation for the CARMENES survey
was to search for planets around low-mass stars; our sample of magnetic field
measurements therefore underrepresents stars more massive than
0.5\,M$_{\odot}$. Most of the stars from \citet{2020A&A...635A.142K} have
masses close to the solar value. The sample covers three orders of magnitude
in $P$ and $Ro$ with a strong concentration around $Ro = 0.6$; 51 stars
(16\,\%) of the sample are in the $0.5 < Ro < 0.8$
($\log{g} \approx -0.2$) range. Although our sample mainly consists of stars that
were observed to discover exoplanets, it was not intentionally biased towards
slowly rotating or inactive stars. The rotational distribution of the sample
is therefore a fair, albeit certainly not unbiased, representation of the
(non-binary) stars in the local Galaxy. The distribution implies systematic
biases and needs to be taken into account for what follows. The same applies
for potential harmonics in the rotational periods, although their impact is
probably small because of the large range in periods covered by our sample.

\subsection{Period-mass diagram}
\label{sect:braking}

\begin{figure}
  \centering
  \resizebox{\hsize}{!}{\includegraphics{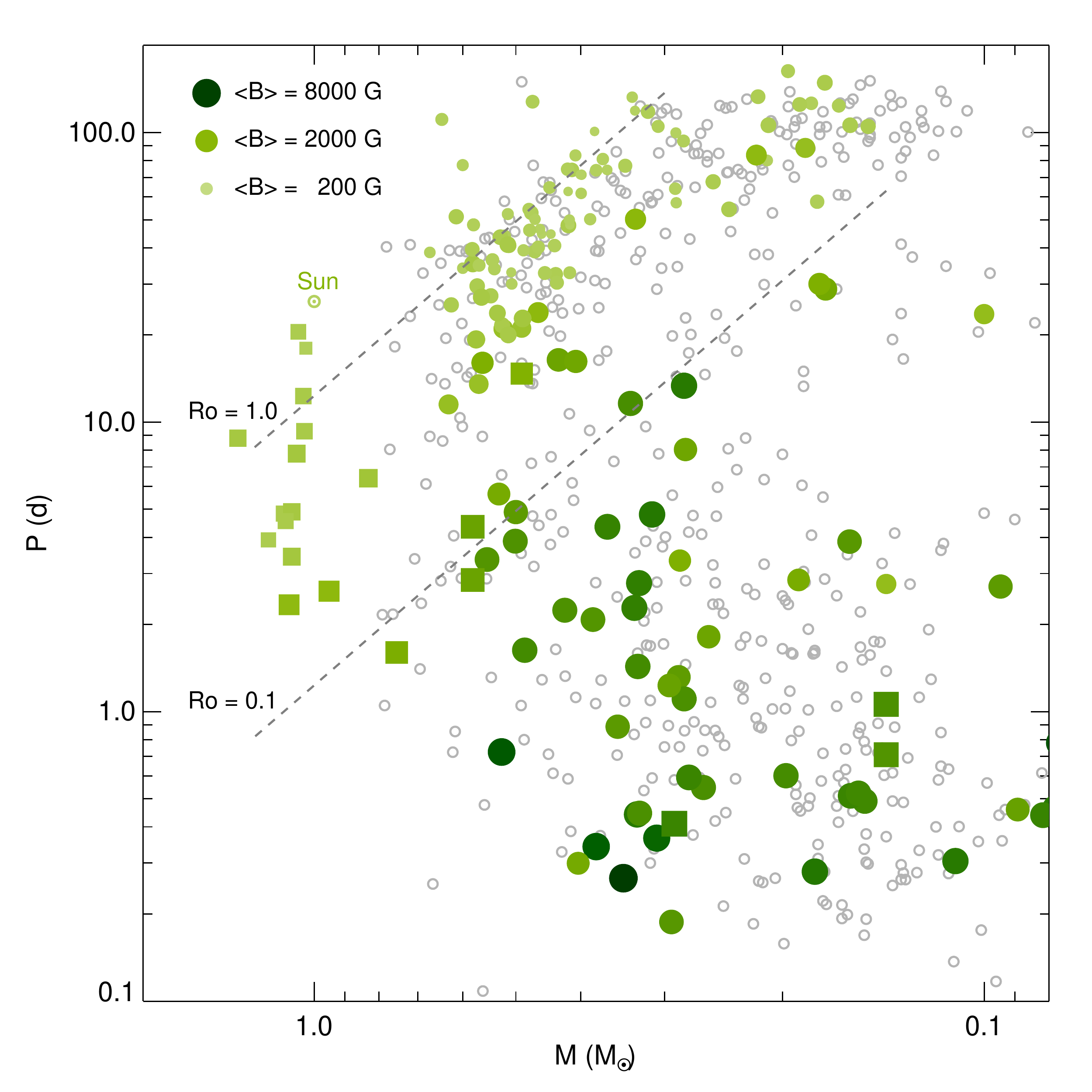}}
  \caption{\label{fig:PM-diagram}Period-mass diagram indicating magnetic field
    strength in color and symbol size. Data from this work are shown as
    circles, and other data are shown as squares.  Literature values are from
    \citet{2020A&A...635A.142K} for young Sun-like stars and for seven
    additional targets from \citet{2017NatAs...1E.184S}. The solar datum from
    \citet{2004Natur.430..326T} is annotated. Gray open circles show stars
    with known rotation periods as reported by
    \citet{2017ApJ...834...85N}. Dashed lines show values of the Rossby number
    $Ro = 0.1$ and 1.0, assuming that bolometric luminosity
    $L_\textrm{bol} \propto M^4$ and $\tau \propto L_\textrm{bol}^{-0.5}$ for
    dwarf stars.}
\end{figure}

The evolution in time of stellar rotation as a function of mass can be
followed in a period-mass diagram \citep[e.g.,][]{2003ApJ...586..464B,
  2011ApJ...727...56I, 2013A&A...560A...4R, 2014ApJS..211...24M}. Observations
of star clusters and field stars show the rotational evolution of stars from
rapid rotators with very short rotational periods ($P < 1$\,d) to slower
rotators with periods of $P = 10$\,d and longer depending on stellar
mass. Stars are believed to lose angular momentum through the interaction
between stellar wind and magnetic fields \citep[e.g.,][]{1968MNRAS.138..359M,
  1970saac.book..385K, 1972ApJ...171..565S, 1981ApJ...248..279P,
  1988ApJ...333..236K, 2012ApJ...746...43R, 2015ApJ...799L..23M,
  2015A&A...577A..98G, 2021LRSP...18....3V}.

Observations of nonthermal energy have shown that stellar activity is
intimately coupled with rotation \citep{1984ApJ...279..763N,
  2003A&A...397..147P, 2011ApJ...743...48W, 2018MNRAS.479.2351W}, and that the
timescales for angular momentum loss and activity reduction depend on stellar
mass \citep{2008AJ....135..785W}. Stellar magnetic fields are the physical
connection between nonthermal emission and rotational evolution. With our
large set of magnetic field observations we can investigate this relation in
detail and across a large parameter range. In
Fig.\,\ref{fig:PM-diagram}, we show the distribution of our sample stars in the
period-mass diagram and indicate the magnetic field strength. We include the
stars with measured rotational periods from \citet{2017ApJ...834...85N} to
show the distribution of available field measurements in context of a larger
sample. We confirm that the two samples appear very similar in the mass-period
diagram, and we refer the reader to \citet{2017ApJ...834...85N} for a discussion of this
distribution in the context of stellar activity and rotational braking.

Figure\,\ref{fig:PM-diagram} can be compared to diagrams visualizing the basic
properties of the large-scale magnetic topologies of cool stars from ZDI, for
example, in \citet[][Fig.\,3,]{2009ARA&A..47..333D} and
\citet[][Fig.\,14]{2021A&ARv..29....1K}. A remarkable feature of such diagrams
is a lack of stars at low masses ($M \la 0.3$\,M$_{\odot}$) and rotation
periods around 10\,d and longer (or $Ro \ga 0.1$). This can partly be
explained by a detection bias: small and slowly rotating stars have low
Doppler broadening below the threshold values for ZDI. A low density of stars
is also visible in Fig.\,\ref{fig:PM-diagram} for masses below
0.3\,$M_{\odot}$ in the period range around 10--40\,d. However, stars with
rotation periods around $P = 100$\,d do appear in this mass range. There is no
obvious reason why rotation periods of several tens of days should be more
difficult to detect than longer ones. Potential explanations for the low
density of stars in this parameter range include enhanced braking efficiency
in the relevant period range \citep[see, e.g.,][]{2016ApJ...821...93N} and
therefore rapid evolution of low-mass stars from a few days to about 100\,d
\citep[similarly to attempts to explain the so-called
\emph{\emph{Vaughan-Preston}} gap;][]{1987A&A...177..131R}, as well as
cancellation of contrast features caused by a transition from dark to bright
surface features \citep[see, e.g.,][]{2019A&A...621A..21R}. The distribution
of magnetic fields in the period-mass diagram shows no peculiar features
beyond a mass-dependent weakening of average magnetic fields with slower
rotation. This is discussed in the following subsection.

\subsection{Rotation-magnetic field relation}
\label{sect:rotact}

\begin{figure}
  \centering
  \resizebox{\hsize}{!}{\includegraphics{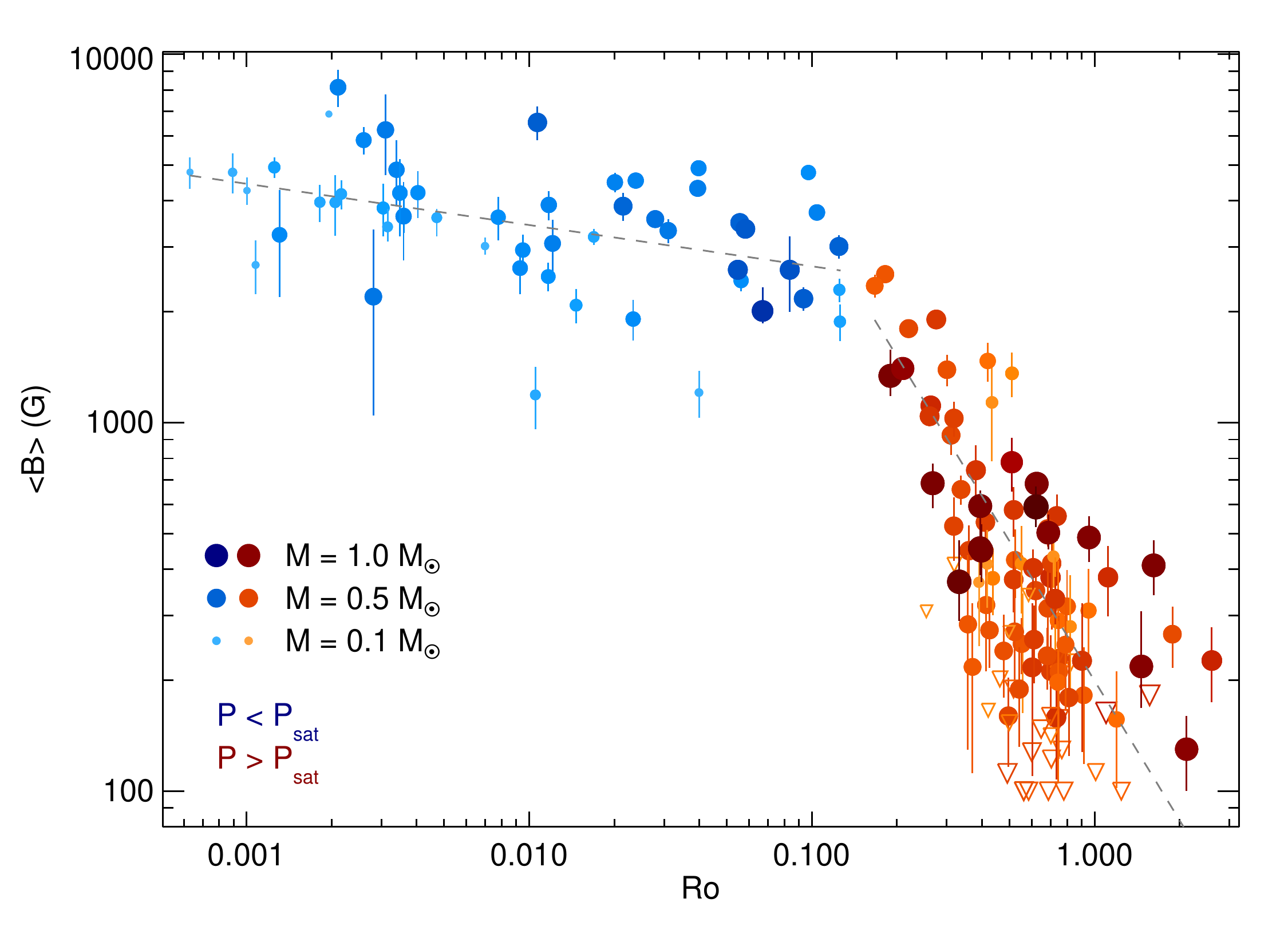}}
  \caption{ \label{fig:BaveRo_diagram} Magnetic field-rotation relation for
    solar-like and low-mass stars. Symbols for stars rotating slower than
    $Ro = 0.13$ are colored red, while those of faster rotators are colored
    blue. Larger and darker symbols indicate higher stellar mass than smaller
    and lighter symbols. The gray dashed lines show linear fits separately for
    the slowly rotating stars ($Ro > 0.13$;
    $\langle B \rangle = 200$\,G\,$\times\, Ro^{-1.25}$) and the fast rotators
    ($Ro < 0.13$; $\langle B \rangle =
    2050$\,G\,$\times\,Ro^{-0.11}$). Downward open triangles show upper limits
    for $\langle B \rangle$.  }
\end{figure}

The decay of average magnetic field strength with rotation coincides with the
well-studied dependence of stellar activity on rotation, as observed in
nonthermal emission. The distribution of magnetic fields in
Fig.\,\ref{fig:PM-diagram} suggests a monotonous relation between rotation and
average magnetic field, which is similar to the rotation-activity relation. The latter
is often expressed as a dependence between normalized chromospheric or coronal
emission from active regions ($L_\textrm{activity}/L_\textrm{bol}$) and the
Rossby number. In Fig.\,\ref{fig:BaveRo_diagram}, we show a similar relation
between the average magnetic field $\langle B \rangle$ and Rossby number.

Our sample reveals a clear dependence between average magnetic field and
Rossby number over more than three orders of magnitude in $Ro$. Similarly to the
rotation-activity relation, the rotation-magnetic field relation exhibits a
break between slow and rapid rotators, the \emph{\emph{saturated}} and the
\emph{\emph{non-saturated}} groups of stars. The non-saturated group have Rossby
numbers above $Ro = 0.13$ (shown as red symbols in
Fig.\,\ref{fig:BaveRo_diagram} and the following figures). In this group, the
average magnetic field strongly depends on $Ro$. In the saturated group (blue
symbols), the average field strength shows a much weaker dependence on
rotation. The rotation-magnetic field relation of the saturated group was
already apparent in Fig.\,3 of \citet{2017NatAs...1E.184S} and Fig.\,12 of
\citet{2021A&ARv..29....1K}.

% table with fits for all masses only
\begin{table}
  \centering
  \caption{\label{tab:rotact}Relations between average magnetic field
    strength, $\langle B \rangle$ (in G), and Rossby number, Ro, and between
    magnetic flux, $\Phi_{\rm B}$ (in Mx) for the slow rotators (and the ratio
    $\langle B \rangle / B_{\rm kin}$ for the fast rotators), and rotation
    period, $P$ (in d).}
    \begin{tabular}{l}
      \hline
      \hline
      \noalign{\smallskip}
      Slow rotation ($Ro > 0.13)$\\
      \noalign{\smallskip}
      \hline
      \noalign{\medskip}
      $\langle B \rangle\,=199\,{\rm G} \times Ro^{-1.26 \pm 0.10}$\\[1ex]
      $\Phi_{\rm B}=5.21\,10^{26}\,{\rm Mx} \times P^{-1.25 \pm 0.07}$\\[1ex]
      \noalign{\smallskip}
      \hline
      \noalign{\smallskip}
      Fast rotation ($Ro < 0.13)$\\
      \noalign{\smallskip}
      \hline
      \noalign{\medskip}
      $\langle B \rangle\,=2050\,{\rm G} \times Ro^{-0.11 \pm 0.03}$\\[1ex]
      $\frac{\langle B \rangle}{B_{\rm kin}}=1.11 \times P^{-0.16 \pm 0.04}$\\[1ex]
      \noalign{\smallskip}
      \hline
      \noalign{\smallskip}
    \end{tabular}
\end{table}

In order to quantify the relation between average magnetic field and Rossby
number, we calculated linear regression curves following the ordinary
least-squares (OLS) bisector method from
\citet{1990ApJ...364..104I}\footnote{\url{http://idlastro.gsfc.nasa.gov/ftp/pro/math/sixlin.pro}}. We
chose the bisector method because the values of $Ro$ come with a relatively
large uncertainty introduced by large systematic uncertainties in the
convective turnover time, $\tau$. Coefficients of the relation are reported in
Table\,\ref{tab:rotact}.  Additionally, our data suggest the existence of two
branches for very slow rotation at $Ro \approx 1$. Stars rotating slower than
this limit ($Ro > 1$) are predominantly partially convective stars (see
Fig.\,\ref{fig:PM-diagram}). Among them, some of the more massive stars' field
strengths seem to depend less on $Ro$ than the overall trend, but this
speculation rests on very few data points only.

An alternative view on the rotation-activity relation is the scaling of
chromospheric or coronal emission (non-normalized instead of normalized) with
rotation period (instead of Rossby number). Such a scaling was suggested by
\citet{1981ApJ...248..279P}, and \citet{2003A&A...397..147P} pointed out that
the convective turnover time approximately scales as
$\tau \propto L_\textrm{bol}^{-1/2}$. This parameterization is in general
agreement with theoretical predictions \citep{1996ApJ...457..340K}, but it is
not obvious to what extent this justifies conclusions about the nature of the
dynamo because $\tau$ likely depends on other parameters as well.
Furthermore, the relevant $\tau$ may exhibit a discontinuity at the fully
convective boundary \citep{2011ApJ...741...54C}, although so far no evidence
for such a discontinuity was found \citep{2018MNRAS.479.2351W}. The scaling of
$\tau$ with $L_\textrm{bol}$ implies that a relation between normalized
emission ($L_\textrm{activity}/L_\textrm{bol}$) and Rossby number ($P/\tau$)
is equivalent to a relation between $L_\textrm{activity}$ and rotation
period. Furthermore, saturation of activity at a fixed Rossby number is
equivalent to saturation at a fixed value of
$L_\textrm{activity}/L_\textrm{bol}$ \citep[see][]{2014ApJ...794..144R}.

\begin{figure}
  \centering
  \resizebox{\hsize}{!}{\includegraphics{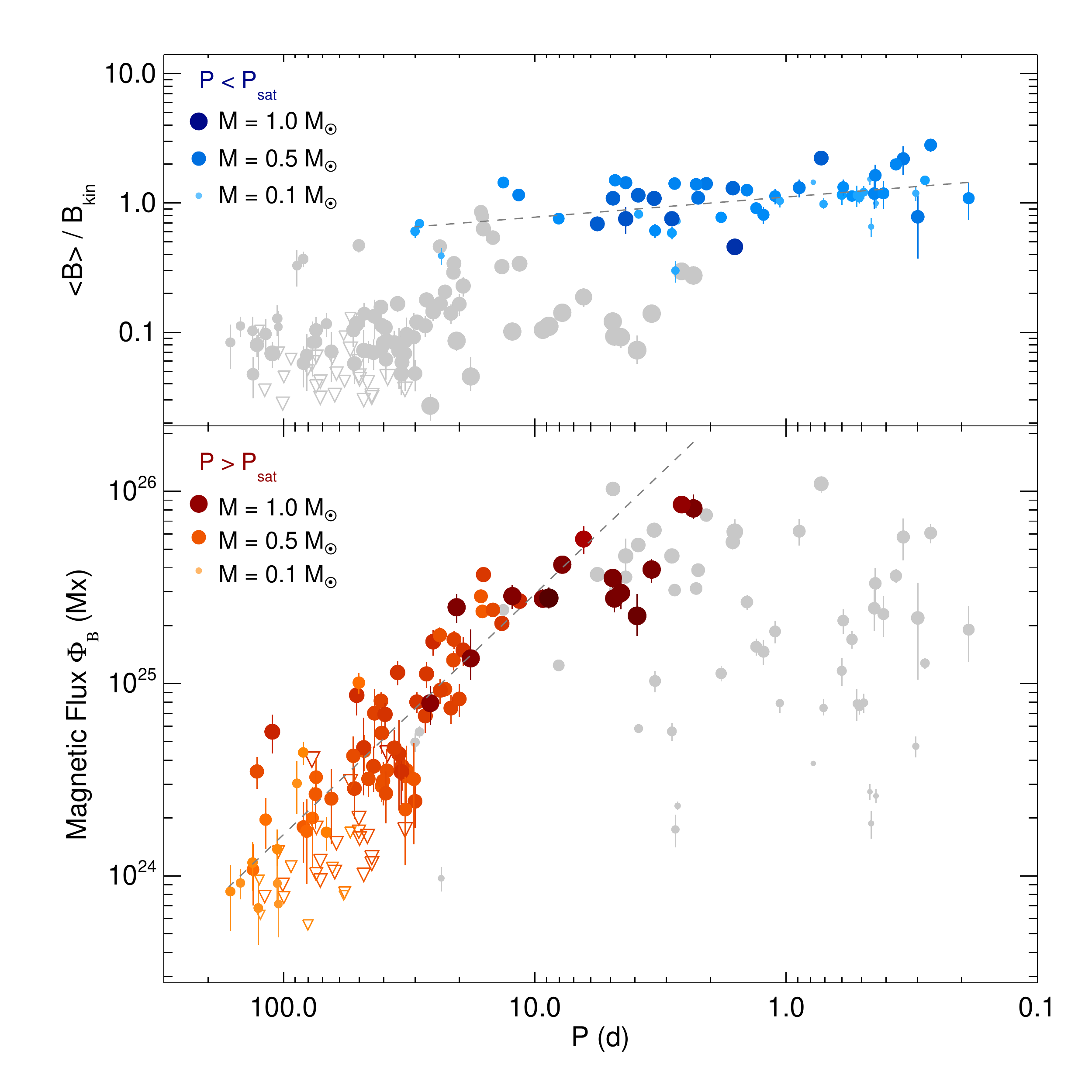}}
  \caption{\label{fig:FluxP_diagram}Alternative version of the
    rotation-magnetic field relation. \emph{Top panel:} Ratio of average
    field, $\langle B \rangle$, to kinetic field limit, $B_\textrm{kin}$, as
    function of rotation period. Stars rotating faster than the saturation
    period are colored blue, while other stars are shown in gray. \emph{Bottom
      panel:} Magnetic flux, $\Phi_\textrm{B}$. Stars rotating slower than the
    saturation limit are colored red, while faster rotators are shown in
    gray. Symbols and colors are the same as in
    Fig.\,\ref{fig:BaveRo_diagram}. See text for details about
    $B_\textrm{kin}$. }
\end{figure}

We investigated scaling laws equivalent to the relation between average
magnetic field and Rossby number (Fig.\,\ref{fig:BaveRo_diagram}) and show an
alternative view on the rotation-magnetic field relation in
Fig.\,\ref{fig:FluxP_diagram}. In its upper panel, we show the ratio between
the average magnetic field, $\langle B \rangle$, and the kinetic field
strength limit \citep{2009ApJ...697..373R},
\begin{equation}
  B_\textrm{kin} = 4800\,\textrm{G} \times
  \left(\frac{ML^2}{R^7}\right)^{1/6}, \label{eq:Bkin}
\end{equation}
with $M$ being the stellar mass, $L$ luminosity, and $R$ radius, all in solar
units. This expression estimates the maximum field strength under the
hypothesis that energy flux determines the magnetic field strength in rapidly
rotating stars \citep{2009Natur.457..167C}. We find that the observed average
field strengths in the rapid rotators indeed populate a relatively narrow
region with values $\langle B \rangle \approx B_\textrm{kin}$. We also find
that the ratio $\langle B \rangle/B_\textrm{kin}$ shows a mild dependence on
rotation with a power law coefficient that is significantly different from
zero (see Table\,\ref{tab:rotact}). Stars rotating slower than the saturation
limit (gray symbols in the upper panel of Fig.\,\ref{fig:FluxP_diagram}) fall
short of this relation. For these non-saturated stars, however, their magnetic
flux, $\Phi_\textrm{B} = 4 \pi R^2 B$, follows a relatively close relation with
rotation period, as is shown in the lower panel of Fig.\,\ref{fig:FluxP_diagram}.

In the non-saturated stars of our sample, magnetic flux shows a clear
dependence on rotational period. We report the relation between
$\Phi_{\textrm{B}}$ and $P$ in Table\,\ref{tab:rotact} and indicate the
relation as a dashed line in the lower panel of
Fig.\,\ref{fig:FluxP_diagram}. A group of stars at $P < 6$\,d shows a somewhat
different behavior, with values of $\Phi_{\textrm{B}}$ significantly below the
overall trend. The group consists of relatively massive stars from the young
Sun sample that may indicate an additional mass- or age-dependence, or that
may be caused by a systematic offset in the literature values. We chose to not
include stars with $M > 0.9$\,M$_{\odot}$ and $P < 6$\,d in our fit (seven
stars). Potential reasons for this deviation from the trend defined by the
lower mass stars include underestimated radii of the young stars (note that
$\Phi_{\textrm{B}} \propto R^2$), an additional dependence of
$\Phi_{\textrm{B}}$ on radius, age, or other parameters, and selection effects
in our sample. We note that for the slow rotators, our relation between
$\langle B \rangle$ and $Ro$ and the one between $\Phi_{\textrm{B}}$ and $P$
are conceptually equivalent because from the equations in
Table\,\ref{tab:rotact}, we estimate
$\langle B \rangle \propto R^{-2} \propto \tau^{1.26}$, and hence
$\tau \propto R^{-1.6}$. For main-sequence stars, this yields approximately
$\tau \propto L^{-0.4}$, which is consistent with the scaling we assumed
between $\tau$ and $L,$ as discussed above.

\subsection{Predictive relations}

\begin{figure}
  \centering
  \resizebox{\hsize}{!}{\includegraphics{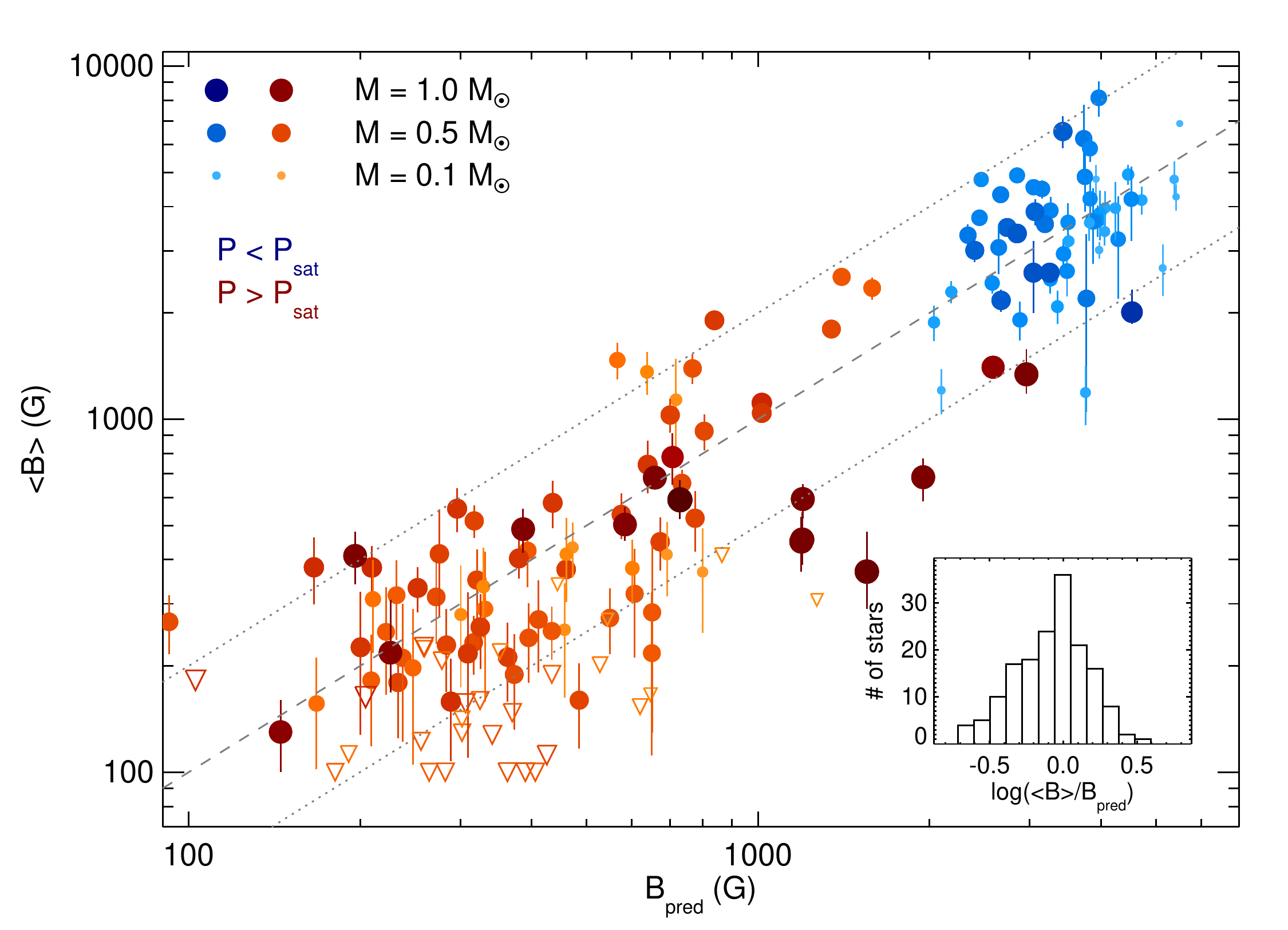}}
  \caption{\label{fig:BPred}Values for predicted value of average
    magnetic field calculated from Eqs.\,\ref{eq:fieldrelations1} and
    \ref{eq:fieldrelations2} valid for all masses. The dashed gray line
    indicates identity between $\langle B \rangle$ and $B_\textrm{pred}$, and
    the histogram shows the distribution of $\langle B \rangle$ around
    $B_\textrm{pred}$. We derive from the histogram that our relation
    estimates the value for $\langle B \rangle$ within roughly a factor of
    two uncertainty; this range is indicated with the two dotted lines.}
\end{figure}

Our results allow us to predict stellar magnetic fields from fundamental stellar
parameters and rotational period, and therefore provide a missing link for
physically consistent models of nonthermal emission
\citep{2017ARA&A..55..159L}, cool star mass loss \citep{2011ApJ...741...54C},
and angular momentum evolution \citep{2015A&A...577A..98G}. One of our main
results is that the average magnetic field of a main-sequence star with
$M \la 1$\,M$_{\odot}$, generated by the magnetic dynamo, can be approximated
from stellar parameters mass, $M$, radius, $R$, luminosity, $L$, (all in solar
units), and rotation period, $P$, (in days) in the following way:
\begin{eqnarray}
  \label{eq:fieldrelations1}
%  B_\textrm{pred} = & 4700\,{\rm G} \times  L^{-0.63} \times P^{-1.26}\quad (\textrm{slow
%  rotation;}\, P > P_{\textrm{sat}}) \label{eq:Bslow}\\
  B_\textrm{pred} = & 8570\,{\rm G} \times  R^{-2} \times P^{-1.25}\quad (\textrm{slow
  rotation;}\, P > P_{\textrm{sat}}), \label{eq:Bslow}\\
  \label{eq:fieldrelations2}
  B_\textrm{pred} = & 5300\,{\rm G} \times \left(\frac{ML^2}{R^7}\right)^{\frac{1}{6}}
  \times P^{-0.16}\quad (\textrm{fast rotation;}\, P < P_{\textrm{sat}}).  \label{eq:Bfast}
\end{eqnarray}
Equation (\ref{eq:Bslow}) follows from the relations between $\Phi_{B}$ and
$P$ for the slow rotators in Table\,\ref{tab:rotact} and is independent of the
choice of $\tau$.  Equation\,(\ref{eq:Bfast}) follows from the relation
between $\langle B \rangle / B_\textrm{kin}$ and $P$ for the fast rotators
together with Eq.\,(\ref{eq:Bkin}). The critical period can be estimated as
$P_\textrm{sat} = 1.6\,\textrm{d} \times
\left(L_\textrm{bol}/L_{\odot}\right)^{-1/2}$, which corresponds to
$Ro = 0.13$. In Fig.\,\ref{fig:BPred}, we show the measured average fields in
relation to the predicted values. In the histogram, we show the distribution
of the ratio $\langle B \rangle / B_\textrm{pred}$. We find that 75\,\% of the
predicted values agree with the measured values within a factor of two.

\subsection{Nonthermal emission}

The magnetic field-rotation relation (Fig.\,\ref{fig:BaveRo_diagram}) shows
remarkable similarity to the activity-rotation relation from X-ray
emission. Other frequently used indicators of stellar activity include the
hydrogen H$\alpha$ line of the Balmer series and the Ca~H\&K lines. X-rays
are emitted at temperatures occurring in the stellar coronae, while both
H$\alpha$ and Ca~H\&K form at lower temperatures in the chromosphere
\citep{1981ApJS...45..635V}. A relation between photospheric magnetic flux,
$\Phi_\textrm{B}$, and X-ray spectral luminosity,
$L_\textrm{X}$, was established by \citet{2003ApJ...598.1387P} that applies to
X-ray irradiance from bright stellar surface regions as well as the total
stellar X-ray output. Such a relation constrains possible heating and emission
models for the Sun and other stars \citep{2016ApJ...830..154F}.

\begin{figure}
  \centering
  \resizebox{\hsize}{!}{\includegraphics{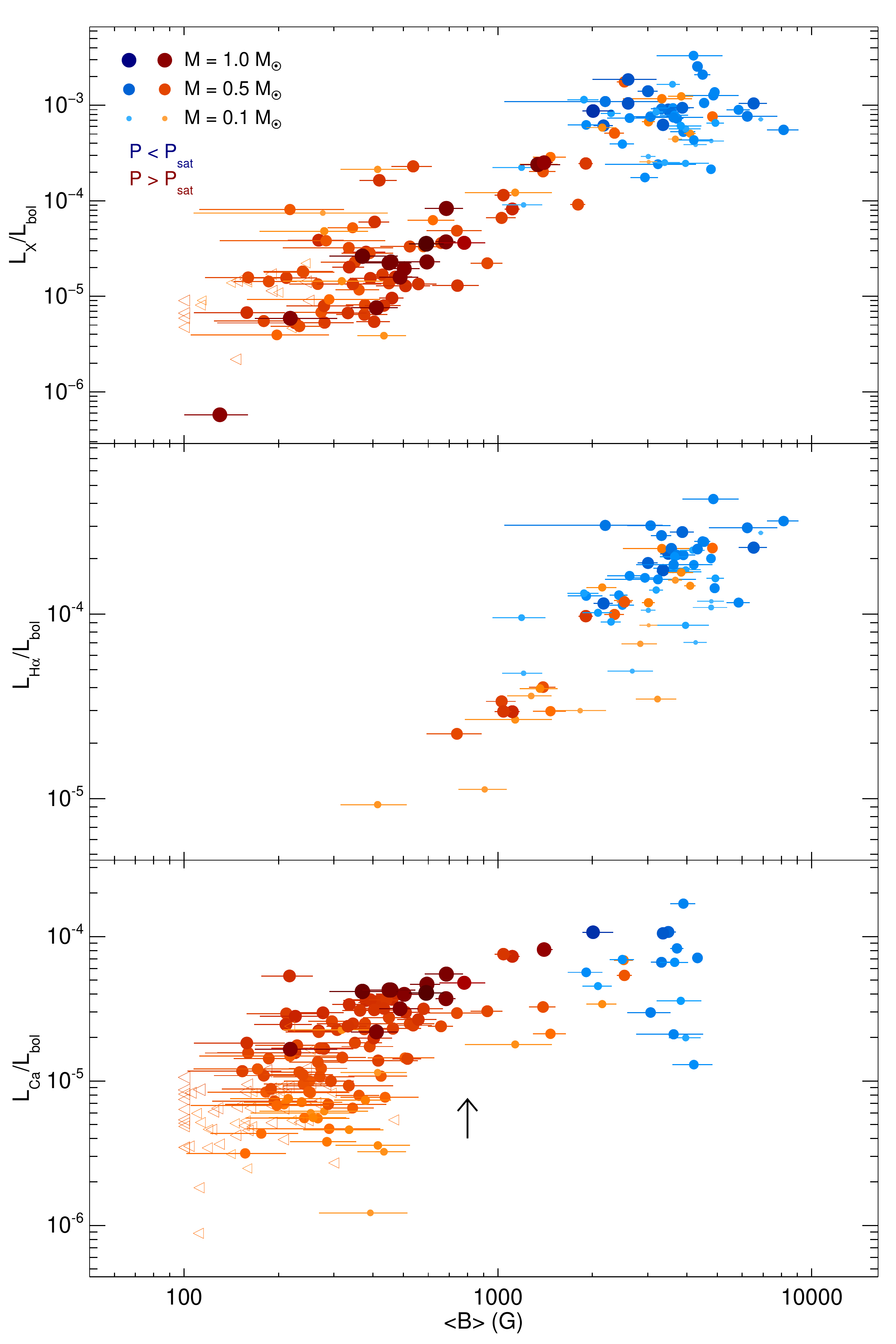}}
  \caption{\label{fig:Emission_B}Normalized coronal (X-ray, \emph{top panel})
    and chromospheric (H$\alpha$, \emph{middle panel}; Ca~H\&K, \emph{bottom
      panel}) emission as a function of average magnetic field. Symbol size
    and color are the same as in Fig.\,\ref{fig:BaveRo_diagram}. The arrow in the bottom
    panel indicates the field strength at which the normalized Ca~H\&K
    emission saturates. }
\end{figure}

Before we turn to the emission/magnetic flux relation, we investigated the
dependence between normalized line luminosity
($L_{\textrm{(X, H}\alpha\textrm{, Ca)}}/L_\textrm{bol}$) and average magnetic
field in Fig.\,\ref{fig:Emission_B}. These relations show a relatively large
scatter but provide information about the typical field strengths required to
generate observable chromospheric and coronal emission. We find that X-ray and
Ca~H\&K emission are observable in stars with very low magnetic field
strengths, which possibly includes a basal component that is unrelated to
stellar magnetic activity \citep{1989ApJ...341.1035S}. On the other hand, we
find that a minimum average field strength of several hundred G is required in
order to generate detectable H$\alpha$ emission in a stellar
chromosphere. This is consistent with chromosphere models in low-mass stars,
showing that Balmer line emission is only generated in the presence of a
sufficiently massive chromosphere \citep{1979ApJ...234..579C}. For Ca~H\&K, we
observe a saturation of the normalized emission at a magnetic field strength
$\langle B \rangle \ga 800$\,G, that is, an increase in the average magnetic
field beyond 800\,G does not lead to an obvious increase in
$L_\textrm{Ca}/L_\textrm{bol}$.

\begin{figure}
  \centering
  \resizebox{\hsize}{!}{\includegraphics{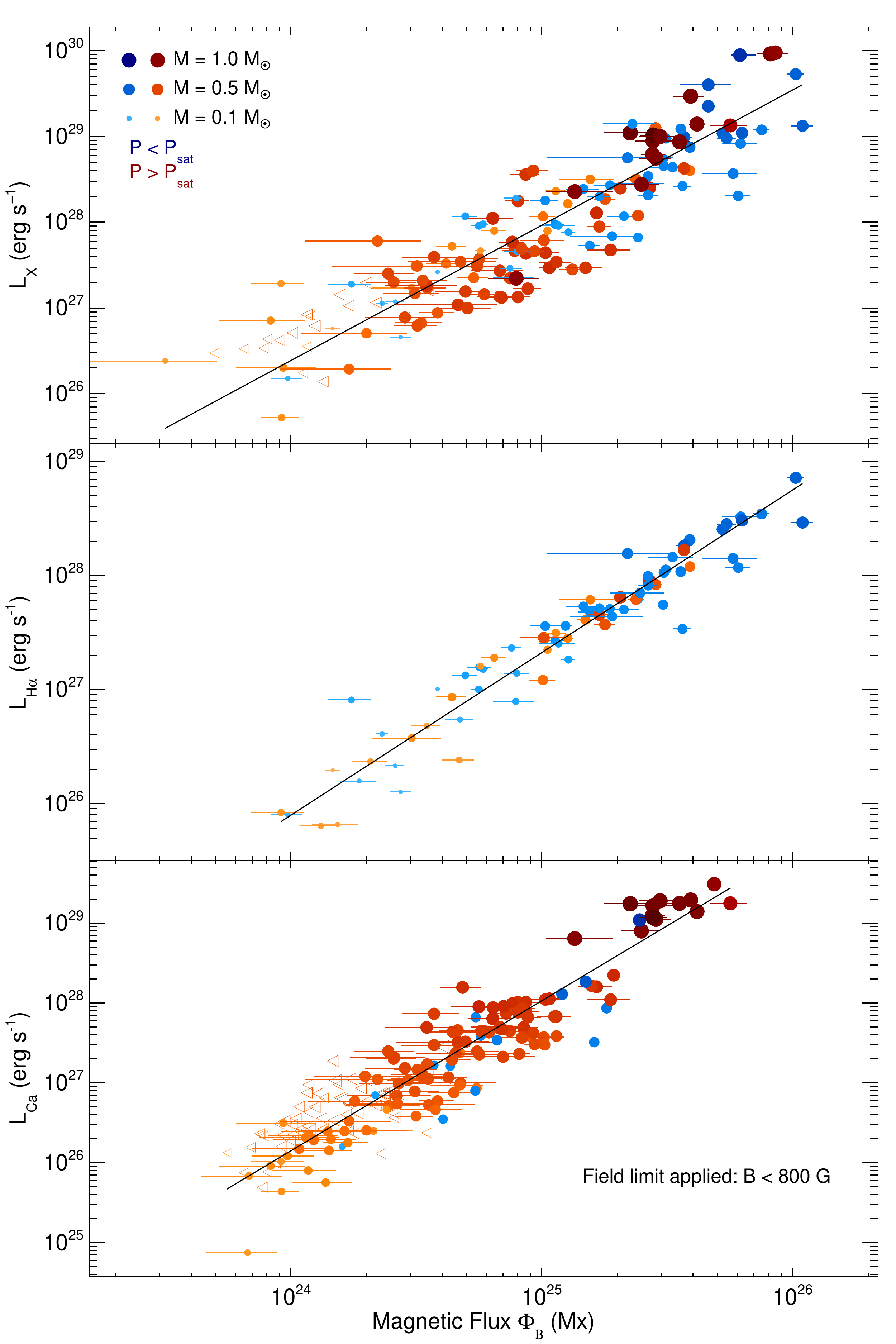}}
  \caption{\label{fig:Emission_Phi}Coronal ($L_\textrm{X}$, \emph{top panel})
    and chromospheric ($L_{\textrm{H}\alpha}$, \emph{middle panel};
    $L_\textrm{Ca}$, \emph{bottom panel}) emission luminosities as a function
    of magnetic flux, $\Phi_\textrm{B}$. Symbol size and color are the same as in
    Fig.\,\ref{fig:BaveRo_diagram}. Linear power laws are reasonable
    approximations for X-ray, H$\alpha$, and Ca luminosities describing both
    rapid and slow rotators. Relations are provided in
    Eqs.\,\ref{eq:emissionrelations1}--\ref{eq:emissionrelations3}. Ca~H\&K
    emission shows a saturation at magnetic flux density
    $\langle B \rangle \approx 800$\,G (see Fig.\,\ref{fig:Emission_B}). This
    limit is applied to the magnetic flux values in the bottom panel
    (Ca~H\&K), but no limit is applied in the other two panels.}
\end{figure}

In Fig.\,\ref{fig:Emission_Phi}, we show that well-defined relations exist
between magnetic flux and emission luminosities in all three stellar activity
indicators, that is, X-rays, hydrogen H$\alpha$, and Ca~H\&K. These relations
show significantly less scatter than those between normalized activity and
average magnetic field in Fig.\,\ref{fig:Emission_B}. Ca~H\&K and X-ray
emission are already visible at magnetic flux levels below
$\Phi_\textrm{B} = 10^{24}$\,Mx. H$\alpha$ emission requires more magnetic
heating and goes together with relatively strong Ca~H\&K emission
\citep{1990ApJS...74..891R}. For the plot showing Ca~H\&K emission (bottom
panel in Fig.\,\ref{fig:Emission_Phi}), we applied a saturation limit of
$B_\textrm{max} = 800$\,G in the calculation of magnetic flux because higher
average fields show no increase in normalized emission for stronger fields
(see above); for all stars with $\langle B \rangle > 800$\,G, we set
$\langle B \rangle = 800$\,G when calculating $\Phi_\textrm{B}$. We interpret
this as saturation of chromospheric Ca~H\&K emission at a field strength of
$\sim$800\,G. We note that the choice of higher maximum field strengths would
essentially shift the blue points (rapid rotators) in the bottom panel of
Fig.\,\ref{fig:Emission_Phi} toward the right; a limit of 1000\,G instead of
800\,G already moves the blue points significantly away from the relation.

We summarize relations between chromospheric and coronal emission and magnetic
flux in Eqs.\,\ref{eq:emissionrelations1}--\ref{eq:emissionrelations3}. The
relations apply to all stars across the entire range of masses and rotation
rates included in our sample (taking into account the field limit in
Ca~H\&K). Our relations quantitatively describe the dependence of nonthermal
chromospheric and coronal emission on the stellar dynamo. With luminosities,
$L$, in erg\,s$^{-1}$ and magnetic flux, $\Phi_{\textrm{B}}$, in Mx, we can write:

\begin{eqnarray}
  \label{eq:emissionrelations1}
  L_\textrm{X} = & 3.28\,10^{-12} \times \Phi_\textrm{B}^{1.58 \pm 0.06}\\
  \label{eq:emissionrelations2}
 L_{\textrm{H}\alpha} = & 4.80\,10^{-9\phantom{1}} \times \Phi_\textrm{B}^{1.43 \pm 0.05}\\
  \label{eq:emissionrelations3}
  L_\textrm{Ca} = & 1.22\,10^{-19} \times \Phi_\textrm{B}^{1.88 \pm 0.05}&
                                                                             (\textrm{apply}
                                                                             B_{\textrm{max}}
                                                                             =
                                                                             800\,\textrm{G})
.\end{eqnarray}

\section{Summary and discussion}

We provided direct measurements of average magnetic field strengths in 292
low-mass main-sequence stars from radiative transfer calculations considering
multiple magnetic field components. For 260 stars of our sample, average field
values are reported here for the first time. Our new data were collected as
part of the CARMENES survey for planets around M dwarfs; in total, we used
15,058 spectra for our analysis, which were corrected for telluric contamination
before co-addition.

The average field strengths we measured span approximately two orders of
magnitude with a lower limit around 100\,G and maximum values of 8000\,G. Not
surprisingly, we observe a relatively large scatter in our investigations of
average field strengths, but a number of clear trends appear that allow us
to draw firm conclusions about the role of average magnetic fields in the
framework of stellar activity. For our analysis, we included literature data
for 22 stars that were obtained with similar methods.

First, we find that the saturation-type, rotation-activity relation, which is
well known from nonthermal coronal emission, can be traced back to a
rotation-magnetic field relation between average field strength, rotation
period, and fundamental stellar parameters. Our data show that rapid and slow
rotators behave differently with a break around $Ro = 0.13$, which is where
the average surface field reaches the kinetic field strength limit. This
demonstrates that a saturation effect in the magnetic dynamo is the reason for
saturation of nonthermal emission instead of a limit in the available stellar
surface area (filling factor equal to unity). We provided relations between
magnetic flux ($\Phi_{\textrm{B}} = 4 \pi R^2 B$) and rotation period valid
for the non-saturated (slowly rotating) stars. From this relation, we derived
a relationship between average magnetic field strength as a function of
rotation period and stellar radius that is independent of the choice of the
convective turnover time. As an equivalent description of the
rotation-magnetic field relation, we also find a relationship between average
magnetic field and Rossby number. We see some systematic deviations between
our data and the predictions, which hints at additional effects that go beyond
our scaling relations. Saturated (rapidly rotating) stars consistently show
average field strengths close to the kinetic field strength limit; the ratio
between average field strength and the kinetic limit is close to unity in the
saturated stars but reveals a mild dependence on rotation.

Second, we investigated relations between nonthermal chromospheric and
coronal emission from X-ray, H$\alpha$, and Ca~H\&K measurements. We observe
that emission luminosity normalized by the stars' bolometric luminosity is
related to average field strengths. In addition, Ca~H\&K line emission shows
saturation at an average field strength around 800\,G. Taking this saturation
into account provides close relations between X-ray, H$\alpha$, and Ca~H\&K
luminosity with magnetic flux for all our sample stars. We reported relations
between magnetic flux and emission luminosities for the three types of
emission lines.

The universal correspondence between magnetic flux and nonthermal emission
sheds some light on the proposed mechanism of centrifugal stripping, according
to which that correspondence was suspected to break down in the most rapid
rotators \citep{2011MNRAS.411.2099J, 2011ApJ...738..164C}. Centrifugal
stripping is consistent with the apparent reduction of flaring activity in
very rapidly rotating M dwarfs with $P \la 0.3$\,d \citep{2020AJ....159...60G,
  2020MNRAS.497.2320R}. A potential mechanism is decreased effective gravity
leading to distortion of magnetic field lines and cooling of the coronal
plasma to chromospheric temperatures \citep{2011ApJ...731..112A}. An
observational signature of this effect would be rapidly rotating stars with
typical chromospheric but abnormally low coronal luminosities with respect to
their magnetic flux. The observed correlation between X-ray emission and
magnetic flux (Fig.\,\ref{fig:Emission_Phi}) shows no evidence for
supersaturation caused by a break in coronal heating. Instead, saturation of
normalized emission with magnetic flux density is visible in the chromospheric
Ca H\&K lines (Fig.\,\ref{fig:Emission_B}). Our results are therefore not
consistent with the coronal stripping scenario.

\begin{figure}
  \centering
  \resizebox{\hsize}{!}{\includegraphics{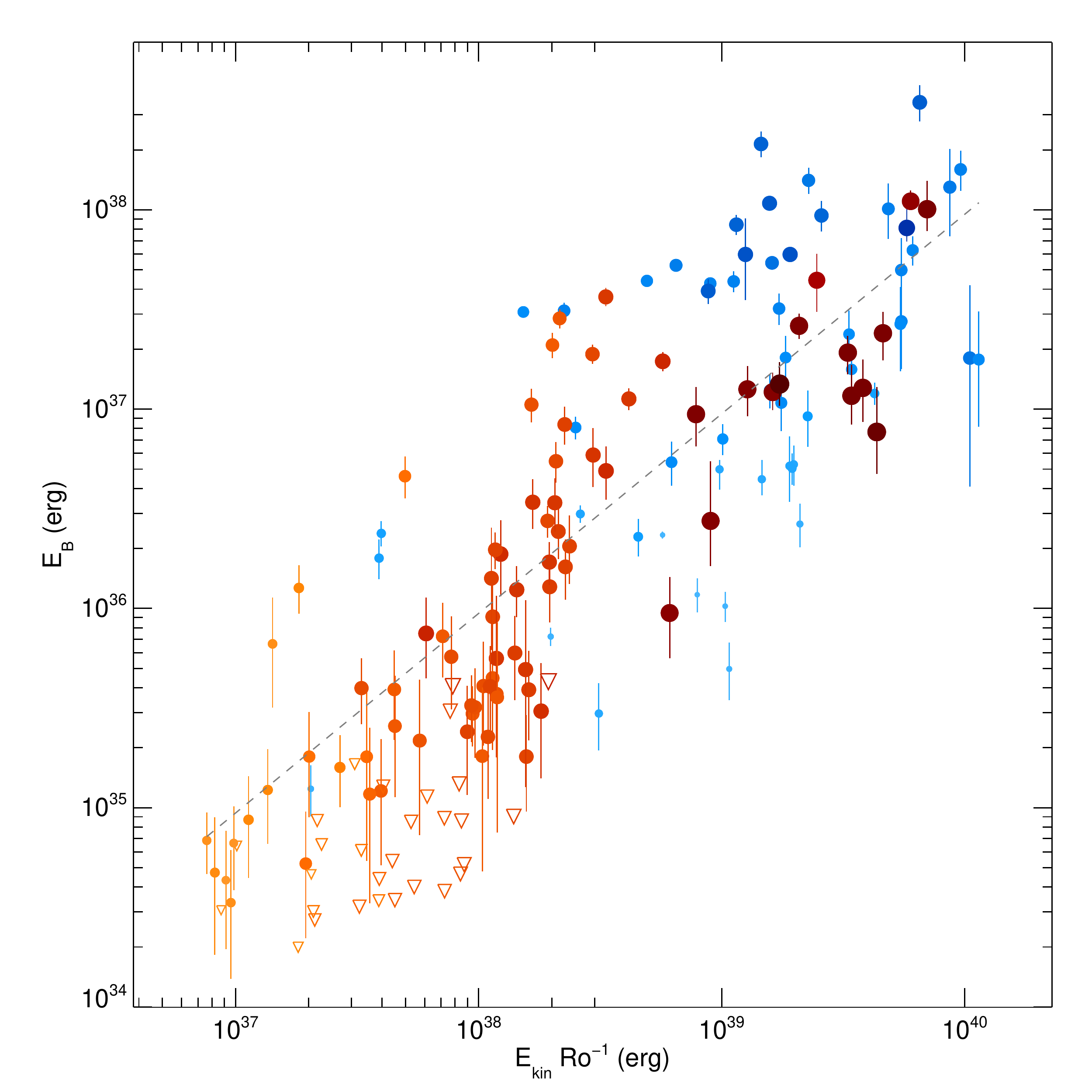}}
  \caption{\label{fig:MAC} Relation between magnetic energy and the ratio
    between kinetic energy and Rossby number for our sample stars. Such a
    relation is motivated by the assumption of a balance between Coriolis,
    buoyancy, and Lorentz forces (MAC balance). Symbols are the same as in
    Fig.\,\ref{fig:BaveRo_diagram}.}
\end{figure}
Our data allow us to test the prediction from force balance (MAC balance) that
magnetic energy grows in proportion to the ratio between kinetic energy and
Rossby number, $E_\textrm{B} \sim E_\textrm{kin}/Ro$
\citep[e.g.,][]{2017LRSP...14....4B}. We show this relation for our sample
stars in Fig.\,\ref{fig:MAC}. A clear correlation is visible, but we can
identify a few obvious trends that show systematic deviations from the
simplified scaling law, for example, at
$E_\textrm{kin}/Ro \approx 10^{39}$\,erg; intermediate mass stars show values
of $E_\textrm{B}$ that are about one order of magnitude larger than the
highest or lowest mass stars in our sample. Our data set provides valuable
input for more detailed tests of dynamo models that go beyond the scope of
this paper.

Looking at all observations together, we conclude that the rotation-activity
relation, including its saturation effect, is a causal consequence of the
characteristics of the magnetic dynamo and its dependence on rotation. We
promote the notion that the magnetic dynamo generates magnetic flux
proportionally to the rotation rate with a limit defined by the available kinetic
energy. Coronal and chromospheric emission are generated with total
nonthermal emission proportionally to magnetic flux. Therefore, nonthermal
emission scales with rotation period until the magnetic field saturation limit
is reached, beyond which point the emission only mildly depends on rotation (because
a mild dependence between magnetic flux and rotation still exists in the
saturation regime). With this background, coronal and chromospheric emission
can be estimated from stellar parameters according to
Eqs.\,\ref{eq:fieldrelations1}--\ref{eq:emissionrelations3}. For example, for
coronal X-ray emission in non-saturated stars, we estimate that
$L_{\textrm{X}} \propto \Phi_{\textrm{B}}^{1.58 \pm 0.06}$ and
$\Phi_{\textrm{B}} \propto P^{-1.25 \pm 0.07}$, which implies
$L_{\textrm{X}} \propto P^{-1.98 \pm 0.07}$, which is consistent with the
observed X-ray activity-rotation relation from much larger samples.

The new observations provide a direct view into magnetic dynamos of low-mass
stars, and they yield a consistent picture of chromospheric and coronal
emission for stars of different masses and rotation periods. The relations
between fundamental stellar parameters, rotation, average magnetic fields, and
nonthermal emission provide useful information for models of stellar and
planetary evolution.

\begin{acknowledgements}
  We thank an anomymous referee for helpful suggestions, and we thank Almudena
  Garc\'ia L\'opez for setting up the electronic data archive.  CARMENES is an
  instrument at the Centro Astron\'omico Hispano-Alem\'an (CAHA) at Calar Alto
  (Almer\'{\i}a, Spain), operated jointly by the Junta de Andaluc\'ia and the
  Instituto de Astrof\'isica de Andaluc\'ia (CSIC). The authors wish to
  express their sincere thanks to all members of the Calar Alto staff for
  their expert support of the instrument and telescope operation.  CARMENES
  was funded by the Max-Planck-Gesellschaft (MPG), the Consejo Superior de
  Investigaciones Cient\'{\i}ficas (CSIC), the Ministerio de Econom\'ia y
  Competitividad (MINECO) and the European Regional Development Fund (ERDF)
  through projects FICTS-2011-02, ICTS-2017-07-CAHA-4, and CAHA16-CE-3978, and
  the members of the CARMENES Consortium (Max-Planck-Institut f\"ur
  Astronomie, Instituto de Astrof\'{\i}sica de Andaluc\'{\i}a,
  Landessternwarte K\"onigstuhl, Institut de Ci\`encies de l'Espai, Institut
  f\"ur Astrophysik G\"ottingen, Universidad Complutense de Madrid,
  Th\"uringer Landessternwarte Tautenburg, Instituto de Astrof\'{\i}sica de
  Canarias, Hamburger Sternwarte, Centro de Astrobiolog\'{\i}a and Centro
  Astron\'omico Hispano-Alem\'an), with additional contributions by the
  MINECO, the Deutsche Forschungsgemeinschaft through the Major Research
  Instrumentation Programme and Research Unit FOR2544 ``Blue Planets around
  Red Stars'', the Klaus Tschira Stiftung, the states of Baden-W\"urttemberg
  and Niedersachsen, by the Junta de Andaluc\'{\i}a. We acknowledge financial
  support from the Agencia Estatal de Investigaci\'on of the Ministerio de
  Ciencia e Innovaci\'on and the ERDF ``A way of making Europe'' through
  project PID2019-109522GB-C5[1:4]/AEI/10.13039/501100011033 % CAB+IAA+IAC+UCM
  and the Centre of Excellence ``Severo Ochoa'' and ``Mar\'ia de Maeztu''
  awards to the Instituto de Astrof\'isica de Canarias (CEX2019-000920-S),
  Instituto de Astrof\'isica de Andaluc\'ia (SEV-2017-0709), and Centro de
  Astrobiolog\'ia (MDM-2017-0737), the Deutsche Forschungsgemeinschaft
  Heisenberg programme (KA4825/4-1), and the Generalitat de Catalunya/CERCA
  programme.
\end{acknowledgements}
  
\bibliographystyle{aa}
\bibliography{refs}

\begin{appendix}

\section{Comparison to \citet{2017MNRAS.472.4563M}}
\label{sect:Moutou}

\begin{figure}
  \centering
  \resizebox{\hsize}{!}{\includegraphics{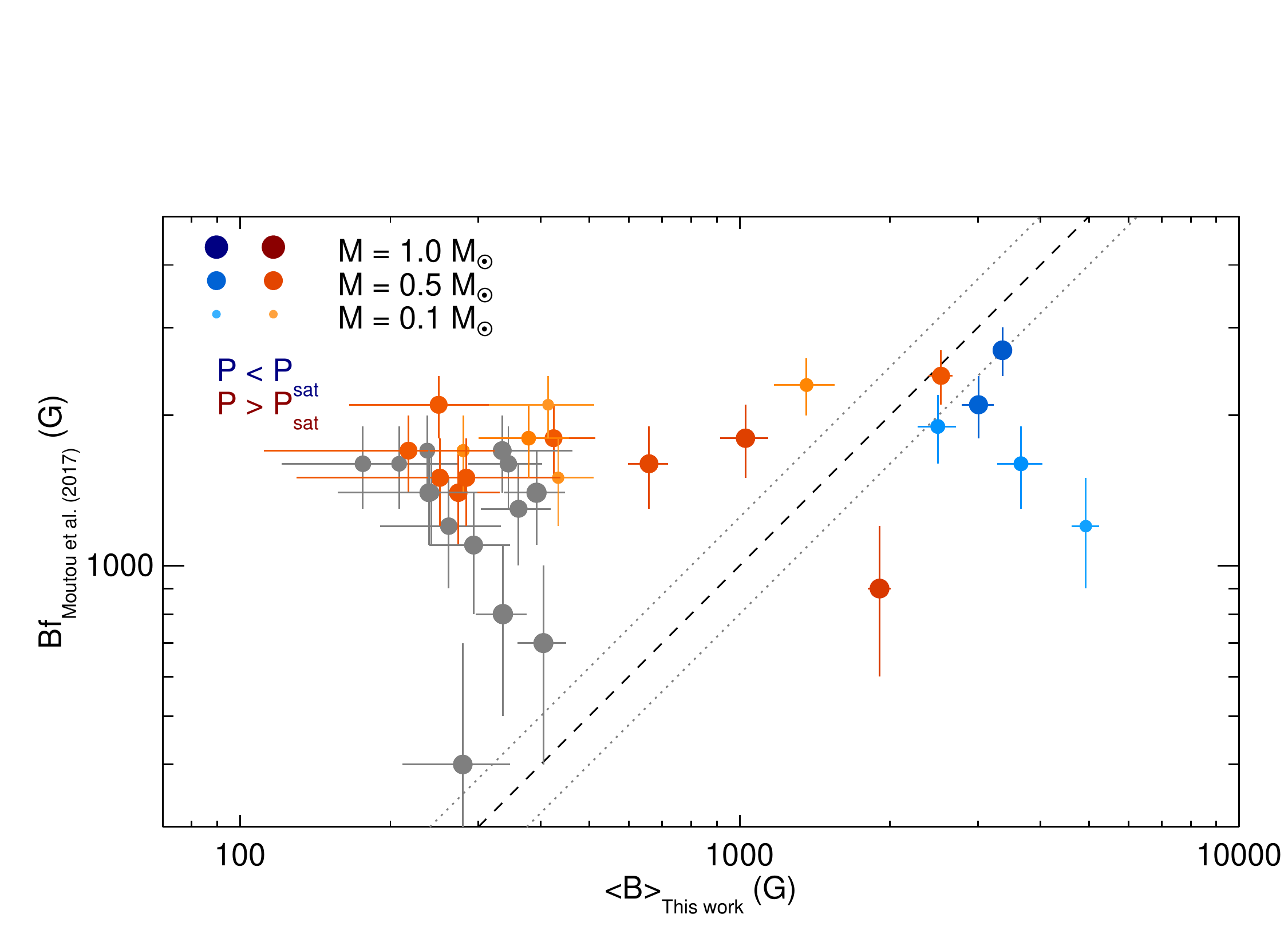}}
  \caption{\label{fig:Moutou}Comparison between average magnetic field
    measurements reported by \citet{2017MNRAS.472.4563M} and our
    results. Lines, symbols and colors are the same as in Fig.\,\ref{fig:Literature}. }
\end{figure}

Similarly to our literature comparison in Fig.\,\ref{fig:Literature}, where we
considered only analyses using multi-component radiative transfer and multiple
lines, Fig.\,\ref{fig:Moutou} shows a comparison between the results from
\citet{2017MNRAS.472.4563M} and our values, extending the comparison carried
out by \citet[][which is shown in their Fig.\,11]{2021A&ARv..29....1K}. Both sets of results show
relatively little correlation. Most of the values from
\citet{2017MNRAS.472.4563M} scatter around $Bf=$\,1--2\,kG, which includes
stars where we measured significantly lower field strengths. A potentially
systematic effect among the slow rotators was already suspected by
\citet{2017MNRAS.472.4563M}.

\section{Star Table}

\longtab[1]{
\scriptsize
   \begin{landscape}
\begin{longtable}{rllcrccccccccccrcc}
       \caption{\label{tab:Sample}Table with stellar parameters and results
         from our analysis}\\
       \hline \hline
       \noalign{\smallskip}
       \# & Karmn & Name & N$_{\rm Obs}$ & S/N & $M$ & $R$ & $T_{\rm eff}$ & $P$ &
       Ref.\,$P$ & $\tau$ & $\log{\frac{L_{\rm bol}}{L_{\odot}}}$ &
       $\log\frac{L_{\rm X}}{L_{\rm bol}}$ & $\log\frac{L_{\rm
           H\alpha}}{L_{\rm bol}}$ & $\log\frac{L_{\rm Ca}}{L_{\rm bol}}$ &
       $\langle B \rangle$ & $\Delta \langle B \rangle$ & $\varv\sin{i}$ \\
       &&&& @8700\,\AA & (M$_{\odot}$) & (R$_{\odot}$) & (K) & (d) && (d) &&&&& (G) & (G)
       & (km\,s$^{-1}$)  \\
       \hline
       \noalign{\smallskip}
       \endfirsthead
       \caption{continued.}\\
       \hline \hline
       \noalign{\smallskip}
       \# & Karm ID & Name & N$_{\rm Obs}$ & S/N & $M$ & $R$ & $T_{\rm eff}$ & $P$ &
       Ref. & $\tau$ & $\log{\frac{L_{\rm bol}}{L_{\odot}}}$ &
       $\log\frac{L_{\rm X}}{L_{\rm bol}}$ & $\log\frac{L_{\rm
           H\alpha}}{L_{\rm bol}}$ & $\log\frac{L_{\rm Ca}}{L_{\rm bol}}$ &
       $\langle B \rangle$ & $\Delta \langle B \rangle$ & $\varv\sin{i}$ \\
       &&&& @8700\,\AA & (M$_{\odot}$) & (R$_{\odot}$) & (K) & (d) && (d) &&&&& (G) & (G)
       & (km\,s$^{-1}$) \\
       \hline
       \noalign{\smallskip}
       \endhead
       \hline
       \endfoot
         1 &          J02565+554W &             Ross 364 &   10 &   270 & 0.61 & 0.61 & 3832 &  51.2 &     SM18 &   45 &    $-$1.14 &        &        &        &      380 &   80 &   0.3 \\
  2 &           J02573+765 &             G 245$-$61 &   54 &   300 & 0.24 & 0.25 & 3381 &     &          &  143 &    $-$2.13 &        &        &        &  $<$ 390 &      &   0.0 \\
  3 &           J03133+047 &               CD Cet &  109 &   660 & 0.18 & 0.21 & 3214 & 126.2 &   New16a &  226 &    $-$2.53 &        &        &    $-$5.22 &      250 &   90 &   0.0 \\
  4 &           J03181+382 &            HD 275122 &   44 &   820 & 0.60 & 0.60 & 3826 &  77.2 &     DA19 &   49 &    $-$1.20 &        &        &    $-$4.93 &  $<$ 360 &      &   1.4 \\
  5 &           J03213+799 &               GJ 133 &   24 &   480 & 0.44 & 0.43 & 3599 &  32.4 &     DA19 &   78 &    $-$1.61 &        &        &        &      320 &  110 &   0.0 \\
  6 &           J03217$-$066 &             G 77$-$046 &   11 &   310 & 0.49 & 0.48 & 3661 &  21.1 &     DA19 &   67 &    $-$1.48 &    $-$4.65 &        &    $-$4.52 &      930 &  110 &   1.4 \\
  7 &           J03463+262 &             HD 23453 &   48 &   780 & 0.60 & 0.60 & 3949 &     &          &   45 &    $-$1.14 &    $-$4.80 &        &    $-$4.40 &      470 &   60 &   1.7 \\
  8 &           J03473$-$019 &             G 80$-$021 &   11 &   280 & 0.50 & 0.50 & 3473 &   3.9 &    Rev20 &   69 &    $-$1.51 &    $-$3.04 &    $-$3.67 &    $-$3.97 &     3490 &  200 &   6.0 \\
\end{longtable}
   \end{landscape}
\normalsize
}

%%% Local Variables:
%%% mode: latex
%%% TeX-master: t
%%% End:

\end{appendix}

\end{document}